\documentclass[%
 reprint,
superscriptaddress,
nofootinbib,
 amsmath,amssymb,
 aps,
]{revtex4-2}
\usepackage{multirow}
\usepackage{graphicx}
\usepackage{dcolumn}
\usepackage{bm}
\usepackage[colorlinks,
            linkcolor=red,
            anchorcolor=blue,
            citecolor=blue
            ]{hyperref}

\graphicspath{ {figure/} }
\begin{document}

\title{Constraints on the spatially dependent cosmic-ray propagation model from Bayesian Analysis}

\author{Meng-Jie Zhao}
 \affiliation{%
 Key Laboratory of Particle Astrophysics, Institute of High Energy Physics, Chinese Academy of Sciences, Beijing 100049, China}
\affiliation{
 University of Chinese Academy of Sciences, Beijing 100049, China 
}%
\author{Kun Fang}
\affiliation{%
 Key Laboratory of Particle Astrophysics, Institute of High Energy Physics, Chinese Academy of Sciences, Beijing 100049, China}
 \author{Xiao-Jun Bi}
\affiliation{%
 Key Laboratory of Particle Astrophysics, Institute of High Energy Physics, Chinese Academy of Sciences, Beijing 100049, China}
\affiliation{
 University of Chinese Academy of Sciences, Beijing 100049, China 
}%



\date{\today}

\begin{abstract}
The energy spectra of primary and secondary cosmic rays (CR) generally harden at several hundreds of GeV, which can be naturally interpreted by propagation effects. We adopt a spatially dependent CR propagation model to fit the spectral hardening, where a slow-diffusion disk (SDD) is assumed near the Galactic plane. We aim to constrain the propagation parameters with the Bayesian parameter estimation based on a Markov chain Monte Carlo sampling algorithm. The latest precise measurements of carbon spectrum and B/C ratio are adopted in the Bayesian analysis. The $\rm{^{10}Be/^{9}Be}$ and Be/B ratios are also included to break parameter degeneracies. The fitting result shows that all the parameters are well constrained. Especially, the thickness of the SDD is limited to 0.4-0.5 kpc above and below the Galactic plane, which could be the best constraint for the slow-diffusion region among similar works. The $\bar{p}/p$ ratio and amplitude of CR anisotropy predicted by the SDD model are consistent with the observations, while the predicted high-energy electron and positron fluxes are slightly and significantly lower than the observations, respectively, indicating the necessity of extra sources.
\end{abstract}
\maketitle


\section{\label{sec:level1}INTRODUCTION}
 The Galactic cosmic-ray (CR) propagation can be described by the diffusion process due to the random scattering by MHD waves in the interstellar medium (ISM). Thus, the properties of the magnetic field turbulence in the ISM determine the CR diffusion. The turbulence in the Galactic disk is mainly generated by stellar feedback (such as the supernova explosions), while in the outer halo, the matter is much rarefied and the turbulence is driven by CRs themselves \cite{Erlykin:2002qg}. As the turbulence origin and ISM properties are both different between the Galactic disk and halo, the CR diffusion in the Galaxy is very likely to be spatially dependent \cite{Evoli:2013lma,Evoli:2018nmb}.

The spatially dependent diffusion is also supported by observations. The TeV gamma-ray halos around some middle-aged pulsars indicate that the diffusion coefficients around these pulsars are more than two orders of magnitude smaller than the average in the Galaxy \cite{HAWC:2017kbo,Aharonian:2021jtz}. If the slow-diffusion zone is common in the ISM around Galactic pulsars, the average diffusion coefficient in the Galactic disk would be significantly suppressed \cite{Hooper:2017gtd}. The spatial magnetic-energy spectrum in the Galaxy also implies that the magnetic field turbulence in the Galactic disk is much stronger than in the halo \cite{doi:10.1146/annurev-astro-091916-055221}, which means that the diffusion coefficient in the Galactic disk could be significantly smaller. Besides, the local CR anisotropy predicted by the standard model \cite{Blasi:2011fm} is much larger than that observed by multiple experiments \cite{Ahlers:2016rox}, while the assumption of a slower CR diffusion in the Galactic disk could give it an explanation \cite{Guo:2015csa}.

The spatially dependent diffusion could account for the well-known spectral hardening of CRs \cite{Feng:2016loc,Guo:2018wyf}. Recent experiments such as PAMELA \cite{Adriani:2011cu}, AMS-02 \cite{Consolandi:2016fhd,Aguilar:2017hno,Aguilar:2018njt}, 
ATIC-2 \cite{Panov:2011ak}, CALET \cite{Adriani:2019aft,Adriani:2020wyg}, DAMPE \cite{An:2019wcw,Alemanno_2021}, CREAM \cite{Yoon:2017qjx} have all discovered the spectral hardening at several hundreds of GeV for most primary and secondary CR nuclei, which cannot be explained by the simplest CR injection and propagation models. Under the spatially dependent diffusion, the energy exponent of the diffusion coefficient can also be spatially dependent, which may explain the spectral hardening. Other possible approaches to interpreting the spectral hardening include the CR injection reflecting nonlinear or time-dependent diffusive-shock-acceleration \cite{Ptuskin:2012qu}, the nonlinear effects in CR propagation \cite{Blasi:2012yr}, and local anomalies due to nearby sources \cite{Thoudam:2011aa} or different transport in the Local Bubble \cite{Ohira:2010eq}.

We expect to use a spatially dependent diffusion model to explain anomalies such as the spectral hardening problem without introducing nearby sources or spectral breaks in the injection spectra. Our model consists of a slow-diffusion disk near the Galactic plane and a fast-diffusion halo more extended in vertical. Variations on diffusion properties can lead to twice spectral hardening for the secondary nuclei compared with the primaries, which is consistent with the AMS-02 observations (see Fig.~84 in \cite{Aguilar:2021tos}). We adopt the Bayesian analysis based on a Markov chain Monte Carlo (MCMC) sampling algorithm to constrain the model parameters, which is meaningful for depicting the CR diffusion pattern in the Galaxy. We also notice that the proton and helium spectra both have a "knee" around 10TV as recently founded by NUCLEON \cite{Gorbunov:2018stf} and DAMPE \cite{An:2019wcw,Alemanno_2021}. We suppose that other mechanisms give this feature and only focus on the spectra below this energy.    

This paper is organized as follows. In Section~\ref{sec:level2}, we introduce our CR propagation model, the data sets used for analysis, and the method of parameter inference. In Section~\ref{sec:result}, we present our fitting results in terms of the parameter posterior probability distributions and the best-fit values. We discuss the parameter constraints by comparing the best-fit spectra and the observations. In Section~\ref{sec:predict}, we adopt our spatially dependent propagation model and the fitting results to predict the $\bar{p}/p$ ratio, the electron and positron spectra, as well as the anisotropy amplitude of CR nuclei, and test if they are consistent with these observations. Section~\ref{sec:conclusion} is the conclusion. 

\section{\label{sec:level2}CALCULATIONS}
\subsection{\label{sec:model}CR propagation model}%
The propagation equation of Galactic CRs is generally expressed by:
\begin{eqnarray}
	 {\frac{\partial \psi}{\partial t}}=&&q(x,p)+\nabla\cdot(D_{xx}\nabla\psi-V_c\psi)
	 +{\frac{\partial}{\partial p}}[p^2D_{pp}{\frac{\partial}{\partial p}}({\frac{\psi}{p^2}})]\nonumber \\\label{eq:trans1}
	 &&-{\frac{\partial}{\partial p}}[\dot p\psi-{\frac{p}{3}}(\nabla\cdot V_c)\psi]
	-{\frac{\psi}{\tau_f}}-{\frac{\psi}{\tau_r}} \label{eq:trans2}
\end{eqnarray}
where $\psi$ is the density of CR particles per unit momentum, $q(x,p)$ is the source term, $D_{xx}$ is the spatial diffusion coefficient, $V_c$ is the convection velocity, $D_{pp}$ is the momentum space diffusion coefficient, $\dot p\equiv dp/dt$ describes ionization and Coulomb losses, $\tau_f$ is the time scales for collisions off gas nuclei, and $\tau_r$ is the time scales for radioactive decay.

Supernova remnants (SNRs) are believed to be the main sources of Galactic CRs, where charged particles are accelerated by shock waves. As suggested by the shock acceleration theory, the injection spectrum of primary CRs is assumed to be a power law as $q\propto R^{-\nu}$, where $R$ is the rigidity of CRs. A low-energy break $R_{\rm br}$ is needed for all the nuclei to fit the observed low-energy spectral bumps. The spectral indices below and above the break are denoted with $\nu_0$ and $\nu_1$, respectively.

The scattering of CR particles on randomly moving MHD waves leads to stochastic acceleration, which is described in the transport equation as diffusion in momentum space $D_{pp}$. Alfv\'en velocity $V_a$ is introduced as a characteristic velocity of weak  propagating in a magnetic field, which is related to the spatial coefficient $D_{xx}$: 
\begin{equation}
	D_{xx}D_{pp}=\frac{4p^2V_a^2}{3\delta(4-\delta)(4-\delta^2)\omega}\,.
\end{equation}
\begin{figure}[htbp]
\includegraphics[width=0.5\textwidth]{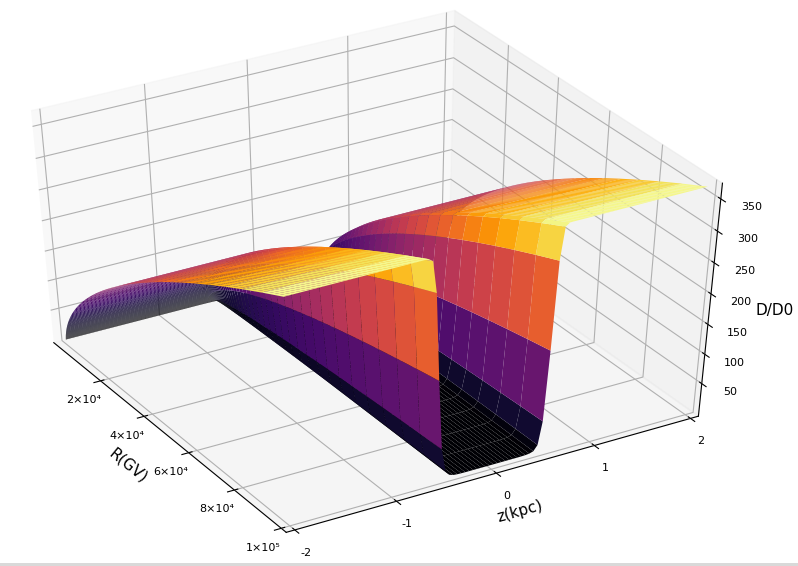}
\caption{The diffusion coefficient $D$ changes with rigidity $R$ and spacial vertical position $z$ where we assume ($N$=8, $h$=0.5~kpc).\label{fig:3d-D}}
\end{figure}

We introduce a slow-diffusion disk (SDD) model, where the diffusion coefficient near the Galactic plane is suppressed. As shown in Fig.~\ref{fig:3d-D}, the SDD model defines the diffusion coefficient $D_{xx}$ by
\begin{subequations}
\begin{equation}
\label{eq:diff}
	D_{xx}(R,z)=aD_0\beta^\eta({\frac{R}{R_0}})^{b\delta}
\end{equation}
\begin{equation}
\label{eq:diff2}
	a=1+(\xi-1){\rm exp}[-({\frac{z}{h}})^N]
\end{equation}
\begin{equation}
\label{eq:diff3}
	b=1+(\xi_{\delta}-1){\rm exp}[-({\frac{z}{h}})^N]\,,
\end{equation}
\end{subequations}
where $\beta=v/c$ is the particle velocity divided by the speed of light, and the low-energy random-walk process is shaped by the factor $\beta^\eta$. Here $\eta\ne1$ is introduced to improve the calculated B/C ratio at low rigidity to fit the observations. The scale factors $a$ and $b$ define the spatial variation of the diffusion coefficient. The scale factor $a$ changes the normalization at the reference rigidity $R_0=4$~GV, while $b$ changes the slope index. The parameter $h$ describes the thickness of this region, and $N$ describes the smoothness of the variation.

We define the dimensionless rigidity parameter $\rho\equiv R/R_0$ so that the diffusion coefficient in the innermost disk ($z\simeq0$, also regarded as local coefficient) can be expressed by $D_i=\xi D_0\rho^{\xi_{\delta}\delta}$. The change of slope index from the halo to the disk can be defined as $\Delta\equiv\delta(1-\xi_{\delta})$. According to a similar spatially dependent propagation model given by Ref.~\cite{Tomassetti:2012ga}, the effective height of the slow-diffusion~(SD) region can be defined as $\Lambda(\rho)\equiv h+(L-h)\xi\rho^{-\Delta}$, where $L$ is the overall size of the Galactic diffusion halo. We can see that the effective height tends to $h$ at high energies.

We assume $N=\infty$ in the main text, which means that the diffusion has a two-zone feature. We discuss the effect of $N$ in Appendix~\ref{app:2}. For the two-zone diffusion scenario, the local CR fluxes and ratios can be approximately expressed by the following forms as given by Ref.~\cite{Tomassetti:2012ga}:
\begin{equation}
\label{eq:pri}
	\psi_{\rm{pri}}(0)\propto \rho^{-\nu}\frac{\Lambda(\rho)}{D_i}=\frac{\rho^{-\nu-\delta}}{D_0}[\frac{h}{\xi}\rho^\Delta+(L-h)]\,,
\end{equation}
\begin{equation}
\label{eq:s/p}
	\frac{\psi_{\rm{sec}}(0)}{\psi_{\rm{pri}}(0)}\propto\frac{\Lambda(\rho)}{D_i}=\frac{\rho^{-\delta}}{D_0}[\frac{h}{\xi}\rho^\Delta+(L-h)]\,,
 \end{equation}
\begin{equation}
\label{eq:u/s}
	\frac{\psi_u(0)}{\psi_s(0)}\propto\frac{\sqrt{D_i\tau_r}}{\Lambda(\rho)}=\frac{\sqrt{\xi D_0\tau_r\rho^{\xi_{\delta}\delta}}}{h+(L-h)\xi\rho^{-\Delta}}\,,
\end{equation}
where $\psi_{\rm{pri}}(0)$ is primary flux, ${\psi_{\rm{sec}}(0)}/{\psi_{\rm{pri}}(0)}$ is secondary to primary flux ratio and ${\psi_u(0)}/{\psi_s(0)}$ is unstable to stable flux ratio.
It can be seen from Eq.~(\ref{eq:pri}) that the primary CR spectra can be described by the superposition of a hard component and a soft component. When the particle rigidity gets larger, the hard component gets dominant, and the spectral index changes from $(\nu+\delta)$ to $(\nu+\delta\xi_{\delta})$. This feature also appears in secondary/primary ratios such as B/C. Unstable/stable ratio such as $\rm{^{10}Be/^{9}Be}$ has a different form related with the decay lifetime $\tau_r$ and can help to break the degeneracy between $\Lambda(\rho)$ and $D_i$. The Be/B ratio is quite complex and shows a similar feature to Eq.~(\ref{eq:s/p}) at high energies and Eq.~(\ref{eq:u/s}) at low energies, which is discussed in detail in Appendix~\ref{app:beb}.

In Appendix~\ref{app:1}, we prove that $\xi_{\delta}$ is required to be very small by the fitting procedure (See Table~\ref{tab:tableh}), which means that the SDD model prefers an energy-independent $D_i$ in the disk. Thus, we fix $\xi_{\delta}=0$ in the main text to simplify the fitting procedure. In this case, the diffusion coefficient in the disk at low energies could be larger than that in the outer halo, which may not be reasonable considering the origin of the ISM turbulence. Thus, we further add a constraint that the former must be always smaller than or equal to the latter, thus the scale factors $\xi_{\delta}$ and $\xi$ are supposed to be equal to 1 below GeV energy.

To solve the propagation equation, we adopt the numerical GALPROP v56\footnote{Current version available at \url{https://galprop.stanford.edu/}} \cite{Strong:1998pw,Strong:1998fr}. The information of the interstellar medium~(gas, radiation and magnetic fields) are considered in GALPROP, which makes the calculated results more realistic. We revise the differencing scheme in the solver by adopting the finite volume method, which is necessary for the spatially dependent diffusion coefficient \cite{Fang:2018qco}.

Solar modulation significantly changes the CR spectra below $\sim$20~GeV. To account for the solar modulation effect, we adopt a simple force-field approximation \cite{Gleeson:1968zza}, where the strength is described by the solar modulation potential $\phi$. According to Table~\ref{tab:exp}, all the AMS-02 and ACE-CRIS (except $\rm{^{10}Be/^{9}Be}$ \cite{2001ApJ...563..768Y}) measurements used in this paper were taken during the same period (2011/05-2016/05), hence we use a uniform $\phi$ to modulate carbon, B/C and Be/B. For $\rm{^{10}Be/^{9}Be}$ ACE-CRIS data \cite{2001ApJ...563..768Y} taken during (1997/08/27-1999/04/09), we use $\phi-(0.1~\rm{GV})$ as an approximation. The $0.1$~GV difference between the two periods is indicated by the long-term observations of the neutron monitor devices \cite{Ghelfi:2016pcv}.

If we use the default values of cross-section given by GALPROP, there is a conflict between old statistics of $\rm{^{10}Be/^{9}Be}$ \cite{2001ApJ...563..768Y,Hams:2004rz} and the newly measured Be/B by AMS-02 \cite{Aguilar:2018njt}, as the former predicts a thin diffusion halo ($\sim3$~kpc) while the latter predicts a thicker one ($\sim6$~kpc) \cite{Evoli:2019iih,DeLaTorreLuque:2021yfq,Weinrich:2020ftb}. Since the current uncertainties of cross-section are quite large ($\sim10-30\%$) \cite{DeLaTorreLuque:2021yfq,Weinrich:2020cmw,Tomassetti:2015nha}, a modification of normalization in the entire energy range $XS$ and the low-energy slope $XS_{\delta}$ of the beryllium production cross-section \cite{Weinrich:2020cmw,Korsmeier:2021brc} can be introduced to reconcile the conflict: 
\begin{equation}
\begin{aligned}
\label{eq:xs}
   \sigma=\sigma^{\rm{default}}\cdot XS\cdot\begin{cases} (\frac{E_{\rm{kin}/n}}{E_{\rm{kin}/n}^{\rm{thresh}}})^{XS_{\delta}}\,,  \quad& E_{\rm{kin}/n}<E_{\rm{kin}/n}^{\rm{thresh}} \\
   1\,,  \quad& \rm{otherwise}\end{cases}\,.
\end{aligned}
\end{equation}
As the cross-section models predict a break around 5~GeV/n energy and a flat behaviour above it, we choose $E_{\rm{kin}/n}^{\rm{thresh}}=5$~GeV/n. 

In summary, the group of free parameters are
\begin{eqnarray*}
    \bm{\theta}=\{D_0,\delta,L,V_a,\eta,\xi,h,A_c,\nu_0,\nu_1,\\
    R_{\rm{br}},\phi,XS,XS_{\delta}\}
\end{eqnarray*}
where $D_0$, $\delta$, $\eta$, $\xi$, and $h$ are the parameters describing the diffusion coefficient, $L$ is the half-width of the total diffusive halo, $V_a$ is Alfv\'en velocity, $A_c$ is the abundance of carbon when fixing the abundance of proton to $1.06*10^6$, $\nu_0$, $\nu_1$, and $R_{\rm{br}}$ are the first and second indices and the break rigidity of overall injection parameters, respectively, $\phi$ is modulation potential for the AMS-02 measurements, and $XS$ and $XS_{\delta}$ are the modification parameters of the beryllium production cross-section.

\subsection{\label{sec:data}Data sets}
According to Ref.~\cite{Johannesson:2016rlh,Schroer:2021ojh,Evoli:2019wwu}, the data group of H-He and heavy nucleons ($Z>2$) have different constraints on propagation and injection parameters. We only use the heavy nuclei data to give a self-consistent constraint on the propagation model. We assume that all the heavy nuclei share the same injection parameters ($\nu_0,\nu_1,R_{\rm{br}}$) and use the carbon flux, $\rm{^{10}Be/^{9}Be}$ ratio, B/C ratio and Be/B ratio to constrain parameters. The latter two ratios are mainly decided by the $\rm(C-N-O)\rightarrow(Be-B)$ series.

Besides the precise measurements of carbon flux, B/C, and Be/B ratios from AMS-02 \cite{Aguilar:2017hno,Aguilar:2018njt}, other data are also included for better parameter constraints, which are listed in Table~\ref{tab:exp}. For the carbon flux, we use the CALET \cite{Adriani:2020wyg}, NUCLEON \cite{Gorbunov:2018stf}, and CREAM-II \cite{Ahn:2009tb} measurements to cover the multi-TeV energy region and the ACE-CRIS measurements \cite{Yuan:2018vgk} to cover the MeV energy region. The low-energy B/C ratio is constrained by the ACE-CRIS data \cite{Yuan:2018vgk}. The data of $\rm{^{10}Be/^{9}Be}$ ratio is taken from ACE-CRIS \cite{2001ApJ...563..768Y} and ISOMAX \cite{Hams:2004rz}. 

\begin{table*}
\caption{\label{tab:exp}Data Used in This Analysis}
\begin{ruledtabular}
\begin{tabular}{cccc}
 Experiment&Energy Range&data points&Reference\\ \hline
 \multicolumn{4}{c}{\textbf{B/C}}\\
 AMS-02(2011/05-2016/05)& 2-2100 GV& 67 &\cite{Aguilar:2018njt}\\
 ACE-CRIS(2011/05-2016/05)& 0.07-0.17 GeV/n& 6 &\cite{Yuan:2018vgk}\\
 \multicolumn{4}{c}{\textbf{Be/B}}\\
 AMS-02(2011/05-2016/05)& 2-2100 GV& 67 &\cite{Aguilar:2018njt}\\
 \multicolumn{4}{c}{$\mathbf{^{10}Be/^{9}Be}$}\\
 ISOMAX(1998/08/04-08/05)& 0.5-1.6 GeV/n& 2 &\cite{Hams:2004rz}\\
 ACE-CRIS(1997/08/27-1999/04/09)& 0.08-0.14 GeV/n& 3 &\cite{2001ApJ...563..768Y}\\
 \multicolumn{4}{c}{\textbf{C}}\\
  NUCLEON(2015/07-2017/06)& 250-17000 GeV/n& 10 &\cite{Gorbunov:2018stf}\\
 CREAM-II(2005/12-2006/01)&  85-7500 GeV/n& 9 &\cite{Ahn:2009tb}\\
  CALET(2015/10-2019/10)*1.27\footnote{a multiplication of 1.27 is described in \cite{Adriani:2020wyg} to get aligned with AMS-02}& 10-1700 GeV/n& 22 &\cite{Adriani:2020wyg}\\
 AMS-02(2011/05-2016/05)& 0.4-1200 GeV/n& 68 &\cite{Aguilar:2017hno}\\
  ACE-CRIS(2011/05-2016/05)& 0.06-0.2 GeV/n& 7 &\cite{Yuan:2018vgk}\\
 Voyager1-HET(2012-2015)&  0.02-0.13 GeV/n& 8 &\cite{Cummings:2016pdr}\\
 \multicolumn{4}{c}{\textbf{B}}\\
 Voyager1-HET(2012-2015)&  0.02-0.11 GeV/n& 8 &\cite{Cummings:2016pdr}\\
 \multicolumn{4}{c}{\textbf{Be}}\\
 Voyager1-HET(2012-2015)&  0.06-0.1 GeV/n& 2 &\cite{Cummings:2016pdr}\\
\end{tabular}
\end{ruledtabular}
\end{table*}

CRs have fully unimpeded access to Voyager 1, free of solar modulation and local interstellar modulation \cite{Strauss_2013,Luo_2015}. Thus, the Voyager 1 data can be regarded as $\phi=0$~GV. We adopt the carbon, boron, and beryllium fluxes of Voyager 1 \cite{Cummings:2016pdr} to break the entanglement between $\phi$ and other parameters.
Electron, positron, and antiproton fluxes are modulated differently, which will be further discussed in Sec.~\ref{sec:e} and \ref{sec:pbar}.

\subsection{\label{sec:mcmc}Bayesian inference and MCMC}%
From the Bayes theorem, the posterior probability distribution of the model parameters is
\begin{equation}
\label{eq:ba2}
P(\bm{\theta}|{\rm D})= \frac{P({\rm D}|\bm{\theta})P(\bm{\theta})}{P({\rm D})}\,,%
\end{equation}
where $\rm{D}$ denotes the used data, $P(\rm{D}|\bm{\theta})=\mathcal{L}(\bm{\theta})$ is the likelihood function, and $P(\bm{\theta})$ is the prior distribution. The quantity $P(\rm{D})$
in the denominator of Eq.~(\ref{eq:ba2}) is the Bayesian evidence, which is a normalizing constant being independent of the model parameters $\bm{\theta}$ and can be neglected in parameter inference. 

MCMC methods are widely used in Bayesian inference and are powerful to sample the high-dimensional parameter space for CR propagation models \cite{Masi:2016uby,Putze:2010zn,Yuan:2017ozr,Johannesson:2016rlh}. We use the public code CosmoMC\footnote{See \url{https://cosmologist.info/cosmomc/}.} as a generic MC sampler to explore parameter space \cite{Lewis:2002ah,Lewis:2013hha}, which uses the Metropolis-Hastings algorithm to generate samples from the posterior distribution \cite{Lewis:2002ah}. It also provides tools for analyzing the posterior distribution and making confidence contour plots \cite{2019arXiv191013970L}.

\section{FITTING RESULTS\label{sec:result}}
\subsection{Posterior distributions of parameters}
\begin{figure*}[htbp]
\includegraphics[width=1.0\textwidth]{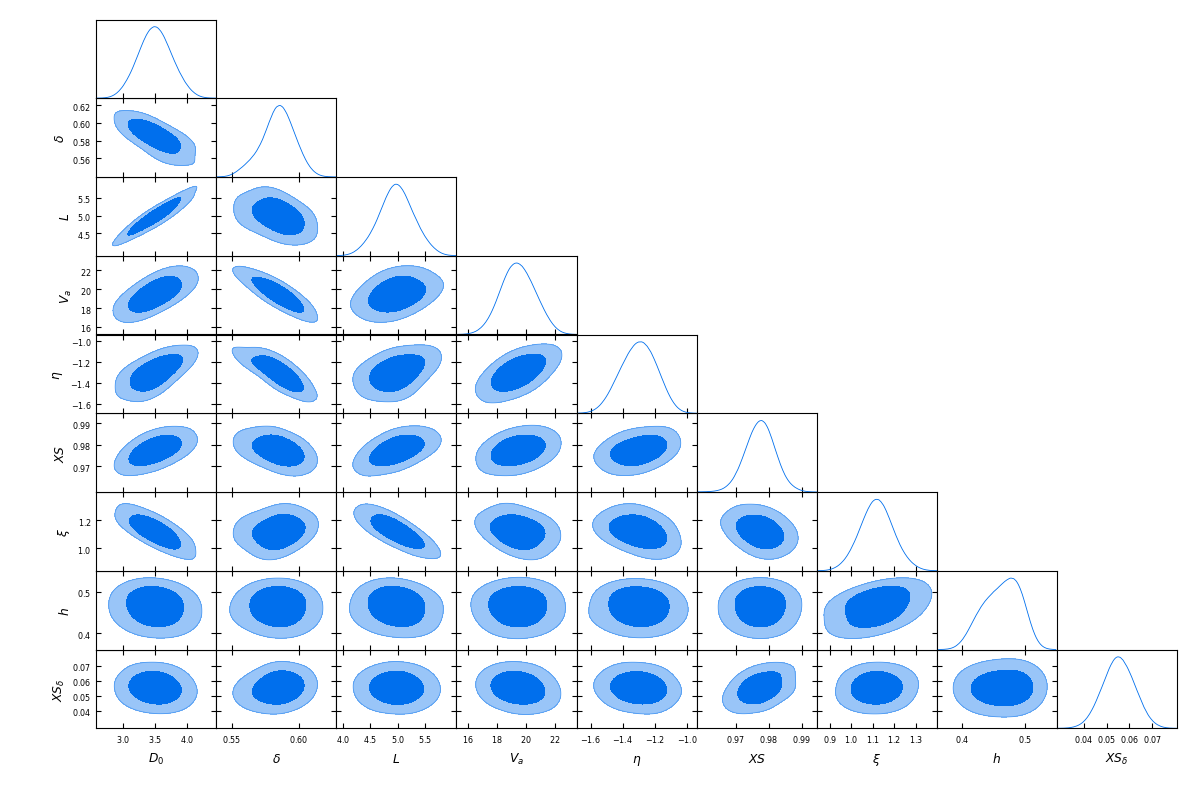}
\caption{\label{fig:MCMC}The 1-D and 2-D distributions of the transport parameters}
\end{figure*}

We first give an expectation about the disentanglement of the important parameters. As mentioned in Section~\ref{sec:model}, the degeneracy between $\Lambda(\rho)$ and $D_i$ can be broken by fitting the data of B/C, $\rm{^{10}Be/^{9}Be}$, and Be/B ratios. Moreover, we have assumed that the diffusion coefficient in the SD must be smaller or equal to that in the halo, which means the scale factors $\xi_{\delta}$ and $\xi$ are equal to 1 at low energies (below $\sim1$~GeV). According to Eq.~(\ref{eq:s/p}) and (\ref{eq:u/s}), the B/C ratio at low energies can be approximated by  
\begin{equation}
\label{eq:s/p2}
	\frac{\psi_{\rm{sec}}(0)}{\psi_{\rm{pri}}(0)}\propto\frac{L}{D_0\rho^\delta}\,,
\end{equation}
while the $\rm{^{10}Be/^{9}Be}$ ratio at low energies can be approximated by
\begin{equation}
\label{eq:u/s2}
	\frac{\psi_u(0)}{\psi_s(0)}\propto\frac{\sqrt{D_0\rho^\delta}}{L}\,.
\end{equation}
Thus, the degeneracy between $D_0$ and $L$ can be broken by the low-energy data combining Eq.~(\ref{eq:s/p2}) and (\ref{eq:u/s2}). According to the definitions of $D_i$ and $\Lambda$, the two parameters $\xi$ and $h$ can be successively determined as long as $D_0$ and $L$ are well constrained. 

Fig.~\ref{fig:MCMC} is the triangle plot of the fitting results, which shows the 1D marginalized posterior probability density functions of the parameters and 2D contour plots of 68\% and 95\% credible regions for all the combinations. The injection parameters are not shown here for simplicity. As we have expected above, all the parameters have well-behaved distributions. Weak anti-correlations in $\delta-V_a$ , $\delta-\eta$, and $\delta-D_0$ can also be seen, which are consistent with the standard models (e.g., Fig.~3 in Ref.~\cite{Johannesson:2016rlh}).

\begin{table*}
\caption{\label{tab:best}The prior range, best-fit values and posterior 95\% range of all parameters in SDD model}
\begin{ruledtabular}
\begin{tabular}{cccc}
 Parameter&Prior range&Best-fit values&posterior 95\% range\\ \hline
 $D_0(10^{28}cm^2s^{-1})$&[0,10.0]&3.379&[2.986,4.023]\\
 $\delta$&[0.2,1.0]&0.583&[0.557,0.608]\\
 $L$(kpc)&[1.0,20.0]&4.743&[4.323,5.625]\\
 $V_a$(km/s)&[0,50]&19.718&[17.130,21.706]\\
 $\eta$&[-3,2]&-1.299&[-1.518,-1.099]\\
 $\xi$&[0,4.5]&1.153&[0.965,1.277]\\
 $h$(kpc)&[0,2.0]&0.468&[0.406,0.515]\\ \hline
 $A_c(10^{3})$\footnote{abundance of proton $A_p$ is $1.06*10^6$, and the normalization of proton flux at 100 GeV is $4.204*10^{-9}\rm{cm}^{-2}\rm{s}^{-1}\rm{sr}^{-1}\rm{MeV}^{-1}$}&[3.1,3.65]&3.337&[3.316,3.377]\\
 $\nu_0$&[0.4,2.0]&1.266&[1.076,1.549]\\
 $\nu_1$&[2.2,2.5]&2.373&[2.364,2.381]\\
 $R_{\rm{br}}$(GV)&[0,5]&1.749&[1.430,2.214]\\
 $\phi$(GV)&[0.5,1.0]&0.782&[0.763,0.793]\\ \hline
 $XS$&[0.7,1.1]&0.973&[0.968,0.986]\\
 $XS_{\delta}$&[-0.2,0.2]&0.0513&[0.0418,0.0689]\\ \hline
 $\chi^2_{\rm{min}}/n_{\rm{dof}}$&-& 167.55/265 &-\\
\end{tabular}
\end{ruledtabular}
\end{table*}

The results are also summarized in Table~\ref{tab:best}, where we list the prior ranges, best-fit values, and posterior 95\% ranges for all the parameters. The total halo height and SDD thickness are well constrained to $L=4.743_{-0.420}^{+0.882}$~kpc, $h=0.468_{-0.062}^{+0.047}$~kpc, respectively.
The halo height $L$ is consistent with the earlier findings using standard models: Ref.~\cite{Weinrich:2020ftb} got $L\in[3.7,6]$~kpc, Ref.~\cite{Putze:2010zn} found $L=4\pm1$~kpc in a pure diffusion/reacceleration model, and Ref.~\cite{Trotta:2010mx} got $L=5.4\pm1.4$~kpc. However, our results are different from another spatially dependent diffusion mode, Ref.~\cite{Feng:2016loc}. They obtained a significantly thicker SDD ($h\sim0.87$~kpc) and larger uncertainty for the halo height, which could be due to the lack of the current precise measurements\footnote{We also notice that they calculated $\chi^2$ by using fewer data on nucleons (above 45~GeV/n) and B/C ratio (above 2~GeV/n), which could also result in a loose constraint.}.

The constrained SDD thickness $h$ is one order of magnitude smaller than the halo height $L$ and a bit thicker than the Galactic disk ($\sim0.2$~kpc) where the CR sources concentrates, which may be explained by the convection of turbulent energy.

The best-fit slope index of diffusion in the halo $\delta=0.583$ is quite larger than the Kolmogorov type (1/3) and a bit larger than the Iroshnikov-Kraichnan type (1/2). The normalization scale factor $\xi$ is around 1, which means that the diffusion coefficient in the disk is close to that in the halo at the reference rigidity of 4~GV.

The constrained modulation potential $\phi\simeq0.782$~GV is in agreement with those found by Ref.~\cite{Yuan:2018vgk}. The two parameters modifying the production cross-section of beryllium, $XS$ and $XS_{\delta}$, are within the experiment uncertainties of $\sigma_{\rm{norm}}\simeq0.2$ and $\sigma_{\rm{slope}}\simeq0.15$ as given in Ref.~ \cite{Weinrich:2020ftb}. 

We note that the best-fit value $\chi^2_{\rm{min}}$ of the SDD model is 167.55 which seems a too good fit compared with the degree of freedom of 265. The reason may be that we have added the systematic errors of these measurements in quadrature with the statistical errors to get the total errors, but have not taken into account the correlation among systematic uncertainties in the calculation. Covariance matrices may be needed to properly take into account those data uncertainties \cite{Derome:2020yfd}.

\subsection{Nucleon fluxes and ratios}
\begin{figure}[htbp]
\includegraphics[width=0.5\textwidth]{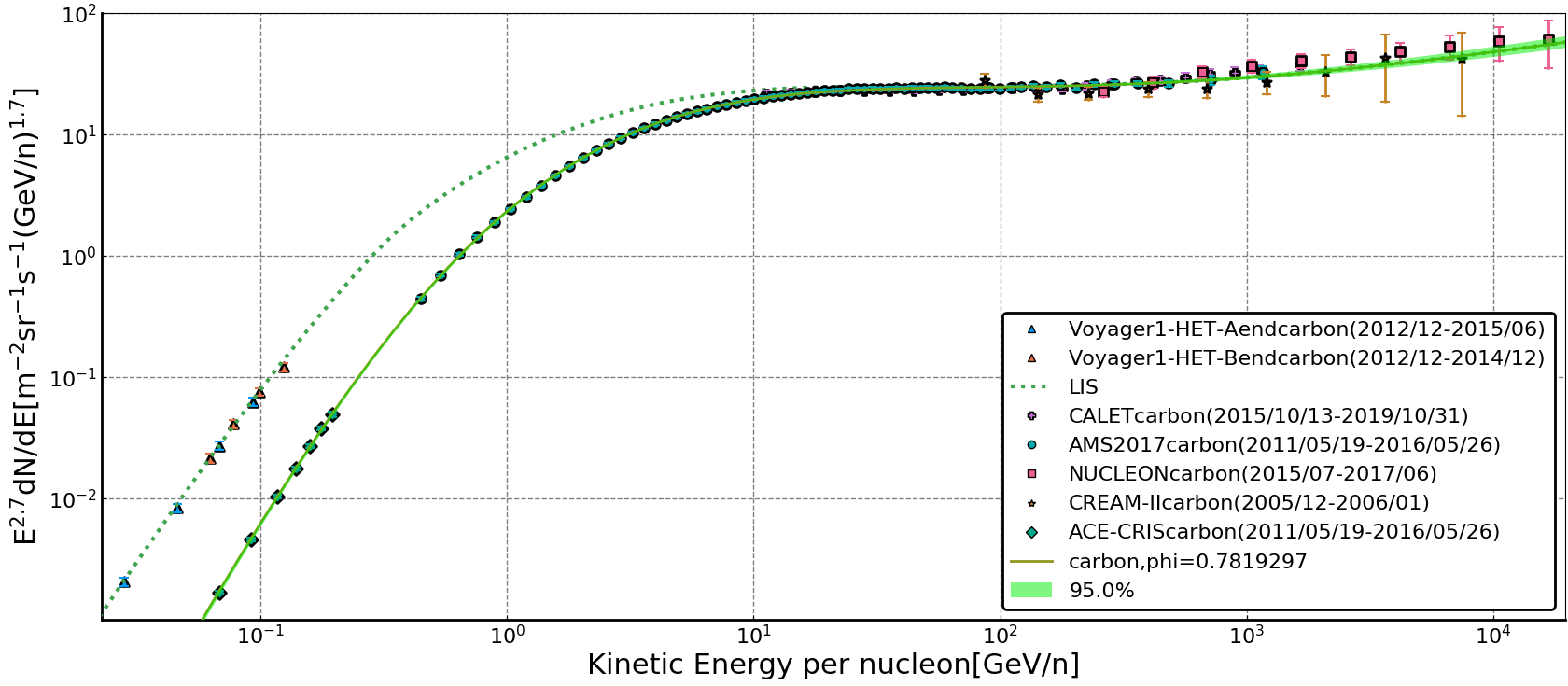}\\
\includegraphics[width=0.5\textwidth]{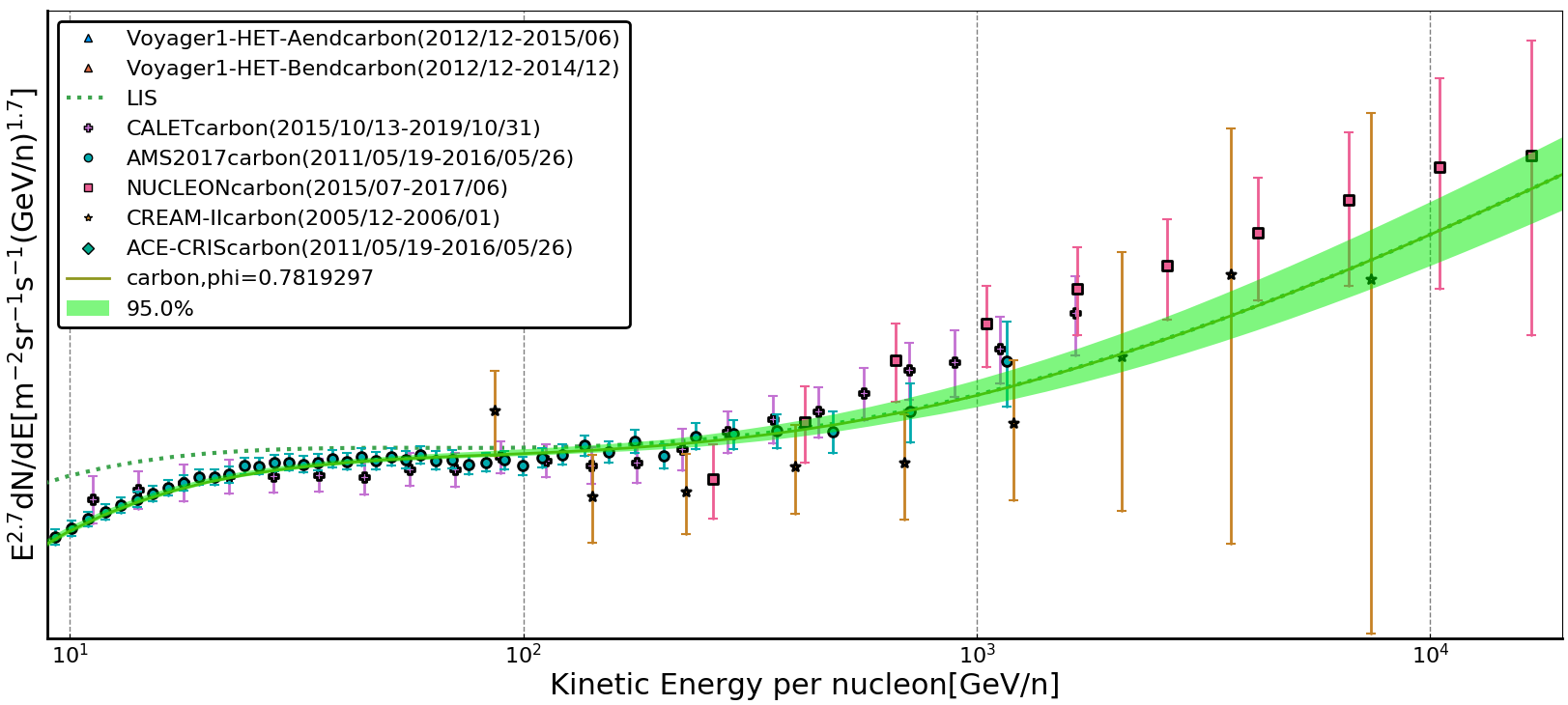}\\
\caption{Carbon spectrum calculated with the best-fit parameters (green solid line) and the $2\sigma$ confidence interval (green band), compared with the experimental data, including AMS-02 \cite{Aguilar:2017hno}, ACE-CRIS \cite{Yuan:2018vgk}, Voyager1 \cite{Cummings:2016pdr}, CALET \cite{Adriani:2020wyg}(multiplied by 1.27), CREAM-II\cite{Ahn:2009tb}, and NUCLEON\cite{Gorbunov:2018stf}. The LIS of carbon are drawn with green dotted line. Top: An wide-range comparison between the model and the experimental data. Bottom: Details around 200~GeV/n where the hardening appears.\label{fig:2sigma-c}}
\end{figure}

The energy spectrum of carbon nuclei is shown in Fig.~\ref{fig:2sigma-c}. A clear spectral hardening can be seen, as predicted by Eq.~(\ref{eq:pri}). At lower energies, the spectrum fits well with the ACE-CRIS measurement, and the LIS also fits well with the Voyager 1 data, which means that the solar modulation potential is reasonable. We have also drawn a 95\% range band derived from the uncertainties of parameters to show that the spectrum is strictly constrained. 

\begin{figure}[htbp]
\includegraphics[width=0.5\textwidth]{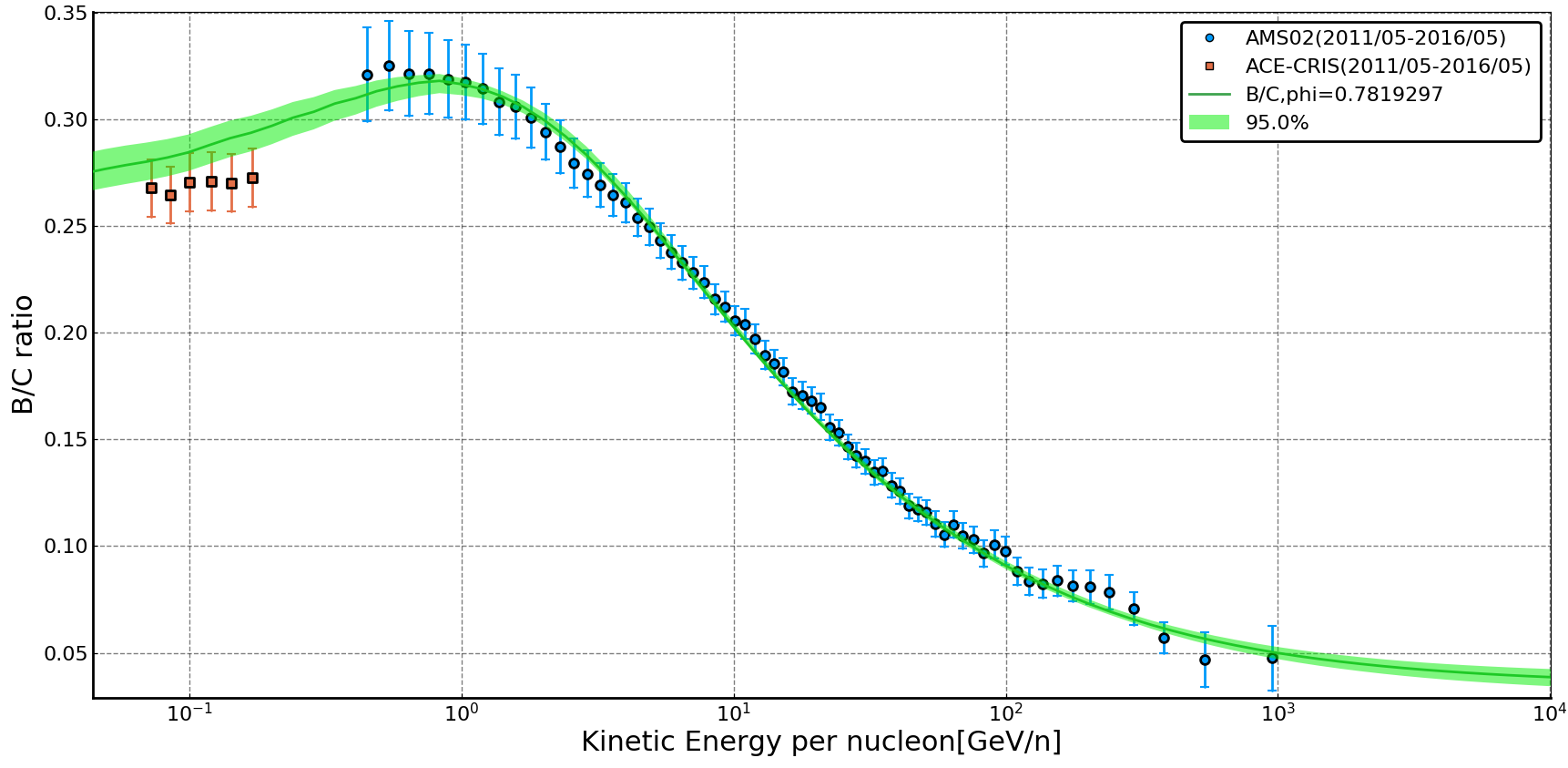}\\
\caption{B/C ratio calculated with the best-fit parameters (green line) and the $2\sigma$ confidence interval, compared with the experimental data: AMS-02 \cite{AMS:2016brs} and ACE-CRIS \cite{Yuan:2018vgk}. \label{fig:2sigma-bc} }
\end{figure}

Eq.~(\ref{eq:s/p}) predicts features of secondary/primary ratios similar to that of the primary spectrum. We can see from Fig.~\ref{fig:2sigma-bc} that our calculation for the B/C ratio fits well with AMS-02 measurement in the entire energy range. The calculation also shows a smooth hardening above 100~GeV/n, which could be confirmed by more precise measurements at higher energies in the future.
At lower energies, the B/C ratio calculated by the SDD model is slightly higher than the ACE-CRIS measurement. We suppose that a modification on boron cross-section or/and solar modulation may explain this difference.

\begin{figure}[htbp]
\includegraphics[width=0.5\textwidth]{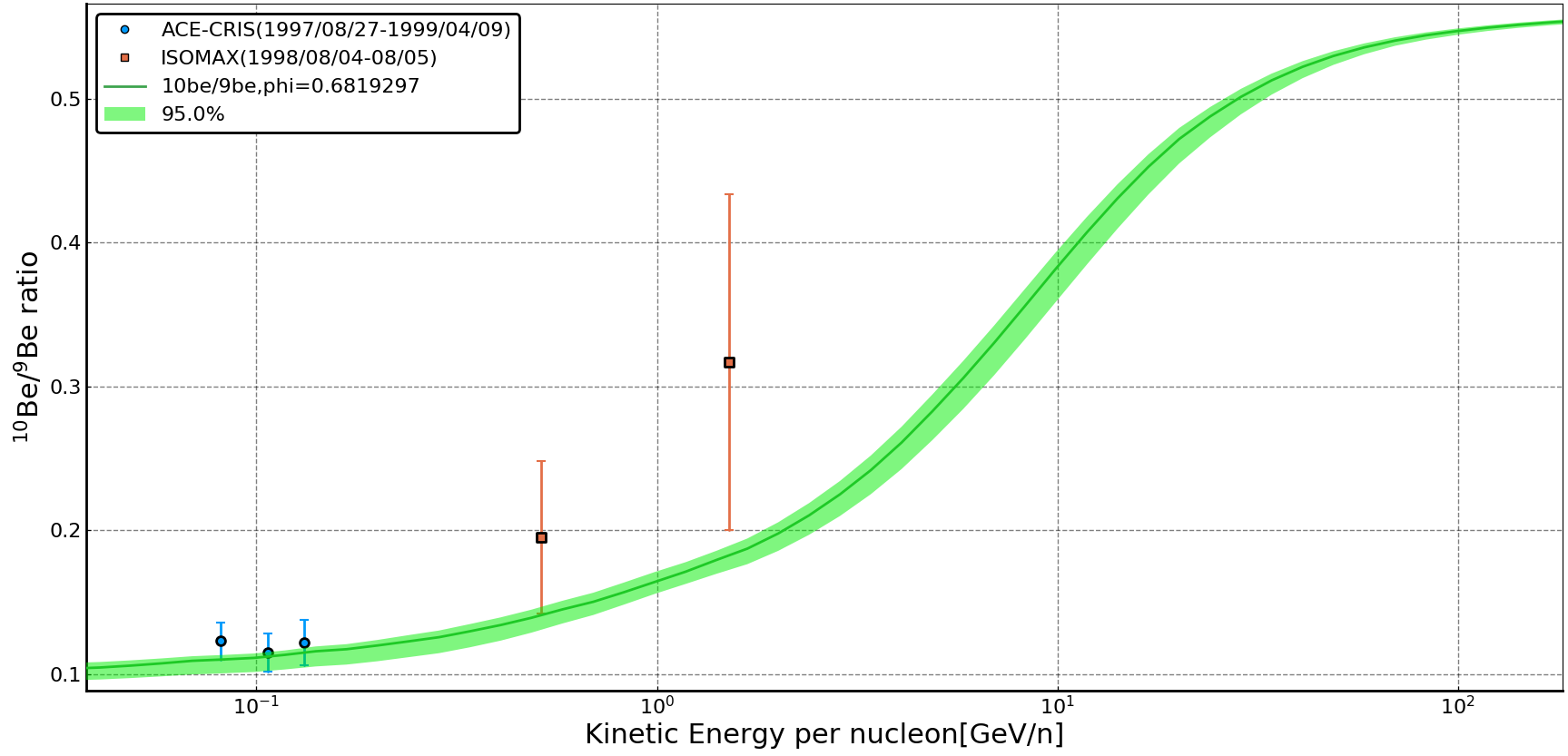}\\
\caption{$\rm{^{10}Be/^{9}Be}$ ratio calculated with the best-fit parameters (green line) and the $2\sigma$ confidence interval (green band), compared with the experimental data: ISOMAX \cite{Hams:2004rz} and ACE-CRIS \cite{2001ApJ...563..768Y}. \label{fig:2sigma-be109} }
\end{figure}

 From Fig.~\ref{fig:2sigma-be109}, the uncertainties reported from ACE-CRIS and ISOMAX
are so large that they cannot give strong constraints on halo height $L$, while more precise experiments on $\rm{^{10}Be/^{9}Be}$ in the future may be required, as well as additional information from Be/B ratio.

From Fig.~\ref{fig:2sigma-beb}, our calculation fits well with the AMS-02 measurement. As analyzed in Appendix~\ref{app:beb}, the Be/B ratio experiences a transition from the decay-dominated low-energy region to the diffusion-dominated high-energy region, which is unique in constraining transport parameters.
According to Ref.~\cite{Weinrich:2020ftb,Tomassetti:2015nha}, the most important region for constraining halo height\footnote{To be exact, effective height $\Lambda$ in SDD model.} should be $10\sim100$~GV where the Be/B ratio has the strongest dependency on it. At energies below 10~GV, the degeneracy with the modification on cross-section and other factors would complicate the fitting.
We also find that the modification of slope $XS_{\delta}$ on the cross-section is needed, otherwise, it is hard to reproduce the feature below 10~GV.

\begin{figure}[htbp]
\includegraphics[width=0.5\textwidth]{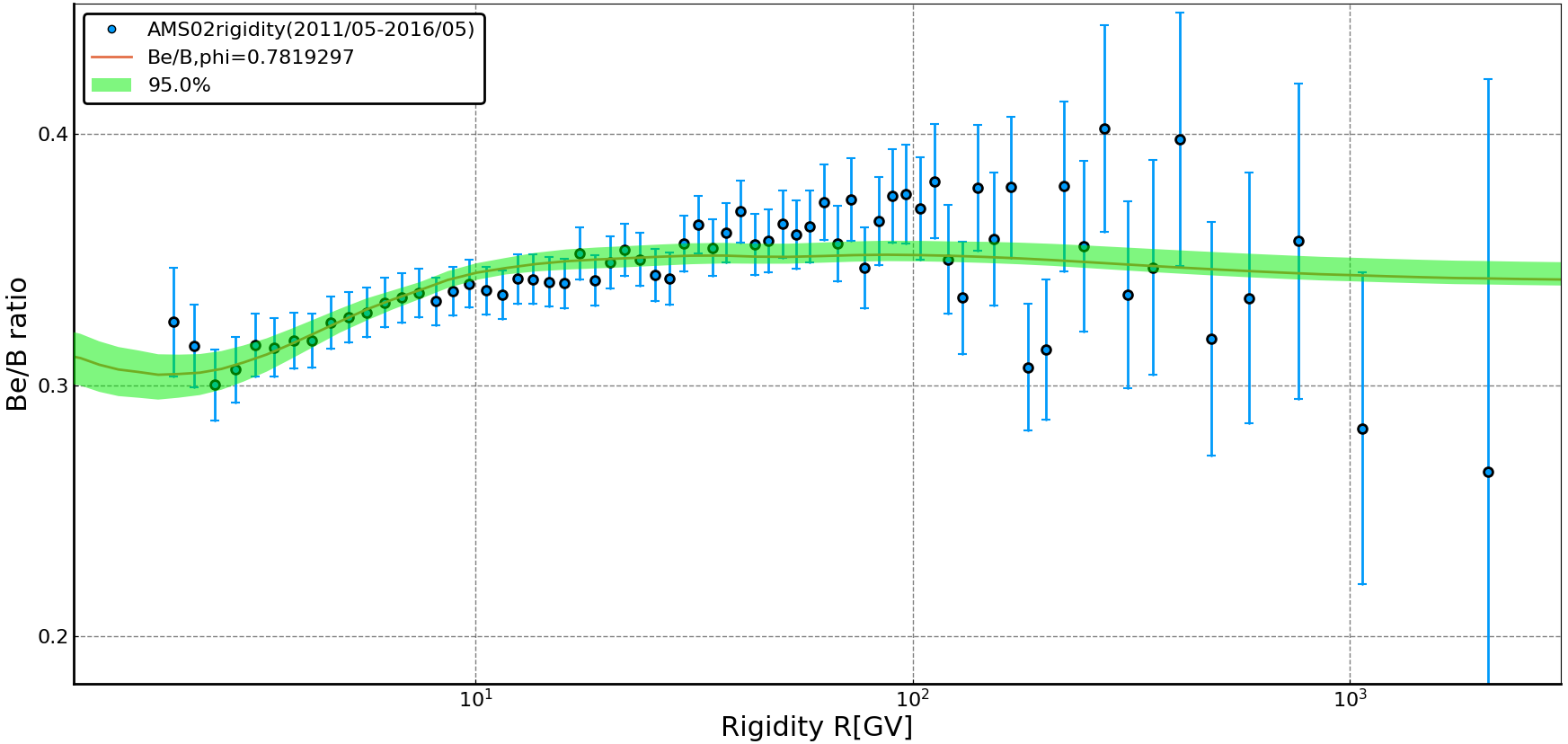}\\
\caption{Be/B ratio calculated with the best-fit parameters (green line) and the $2\sigma$ confidence interval (green band), compared with the experimental data AMS-02\cite{Aguilar:2018njt}. \label{fig:2sigma-beb}}
\end{figure}

\section{PREDICTIONS\label{sec:predict}}
AMS-02 has provided observations on antimatter particles like antiprotons and positrons, which are crucial for predicting dark matter (DM) particles \cite{Aguilar:2019owu,AMS:2016oqu}. The antiproton-to-proton ratio \cite{Giesen:2015ufa} and the positron fraction \cite{Moskalenko:1997gh} predicted by the standard models are significantly lower than the AMS-02 measurements at high energies. The excesses could be explained by introducing DM, while we need to get a proper assessment of antiparticle background firstly. The spatially dependent propagation model may generate higher antiparticle fluxes compared with the standard model \cite{Feng:2016loc}.

Besides, the CR anisotropy predicted by the standard propagation models is significantly higher than the observations. As the anisotropy is proportional to the local diffusion coefficient, our SDD model can effectively suppress the CR anisotropy.

\subsection{Antiprotons\label{sec:pbar}}
To predict the antiproton flux, or $\bar p/p$ ratio for the SDD model, we keep all the parameters in Table~\ref{tab:best} fixed to the best-fit values and change the injection parameters of proton and helium to fit the latest measurements of AMS-02 \cite{Aguilar:2021tos}. We use the default nuclear scaling routine given by GALPROP to get the hadronic cross-sections and calculate antiproton flux.

The production mechanism of secondary nucleons like beryllium is different from that of antiprotons. The nuclear fragmentation keeps the energy per nucleon of secondary particles the same as that of primary particles, while the antiproton spectrum is the convolution of the interstellar spectra and the differential cross-section \cite{diMauro:2014zea}, leading to lower energies compared with the primary particles ($E_{\bar p,\rm{max}}\sim\sqrt{E_p/2}$). So the hardening energy of beryllium and antiproton should be around $\sim200$~GV and $\sim10$~GV respectively.

\begin{figure}[htbp]
\includegraphics[width=0.5\textwidth]{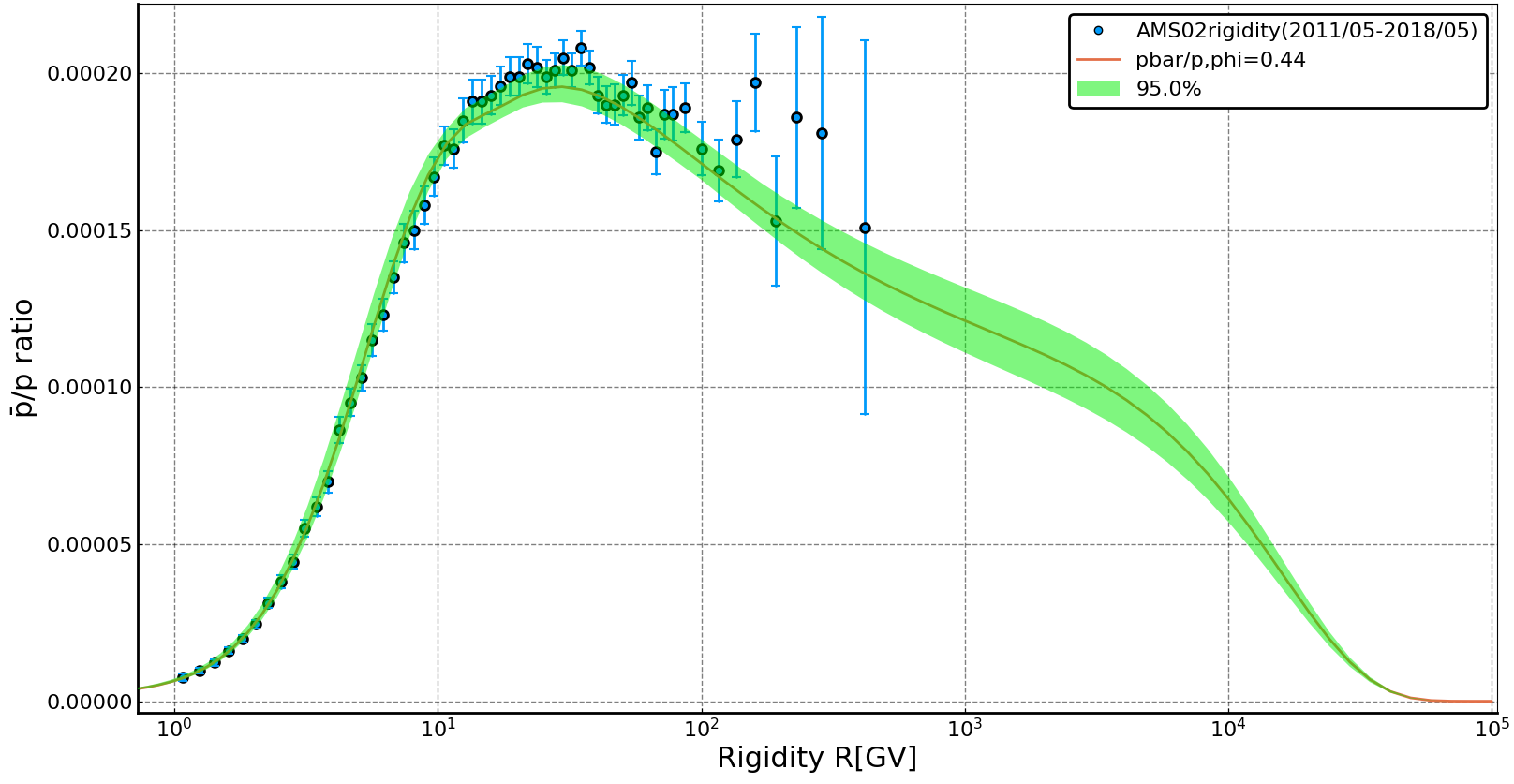}
\caption{$\bar p/p$ ratio predicted by the best-fit parameters obtained in Sec.~\ref{sec:result} (green line) and the $2\sigma$ confidence interval (green band), compared with the experimental data of AMS-02 \cite{Aguilar:2021tos}. \label{fig:2sigma-pbarp} }
\end{figure}

The result in Fig.~\ref{fig:2sigma-pbarp} shows that the SDD model can give a good explanation to the measured $\bar p/p$ ratio without introducing an extra source for antiproton, such as the dark matter annihilation. The predicted hardening above 10~GV by the propagation effect can explain the antiproton excess. Furthermore, the injection and cross-section uncertainties may further improve the fitting result \cite{Giesen:2015ufa}. Here we use a smaller modulation potential $\phi_{\bar p}=0.44$~GV to modulate the low-energy region of antiproton flux, as the charge of antiproton is opposite from proton and should be effected differently by solar activities \cite{Cholis:2015gna}. 
 
\subsection{Electrons and Positrons\label{sec:e}}
As AMS-02 collaboration showed \cite{Aguilar:2019owu,Aguilar:2019ksn} that the
electron and positron spectra each could be well described by the sum of two components. Here we use the SDD model to test if the extra components could be originated from spatially dependent diffusion.

To calculate electrons and positrons fluxes in the SDD model, we reuse the fitted proton and helium fluxes from Section~\ref{sec:pbar} and choose the Pshirkov-ASS model \cite{Pshirkov:2011um} in GALPROP to describe the Galactic magnetic field, which plays a crucial role in lepton energy losses. It has been shown \cite{Evoli:2020szd,2020arXiv201013825D} that SNRs could contribute primary electrons, while a few secondary electrons and positrons can be produced from the decay of charged pions and kaons created in collisions of cosmic-ray particles with gas. Besides, electron-positron pairs can also be produced by PWNe or DM annihilation/decay. To subtract the possible contributions from PWNe, dark matter and secondary electrons, we calculate the primary electrons by using a subtracted form $(\Phi_{e^-}-\Phi_{e^+})$ between electron and positron fluxes to fit the injection of electron \cite{Jin:2015caa}, where we use much recent AMS-02 data \cite{Aguilar:2021tos}. Since the charge of the electron is opposite from proton and $e^+e^-$ have smaller masses than that of the proton, we consider a different $\phi$ to describe how 
they are affected by solar modulation.

\begin{figure}[htbp]
\includegraphics[width=0.5\textwidth]{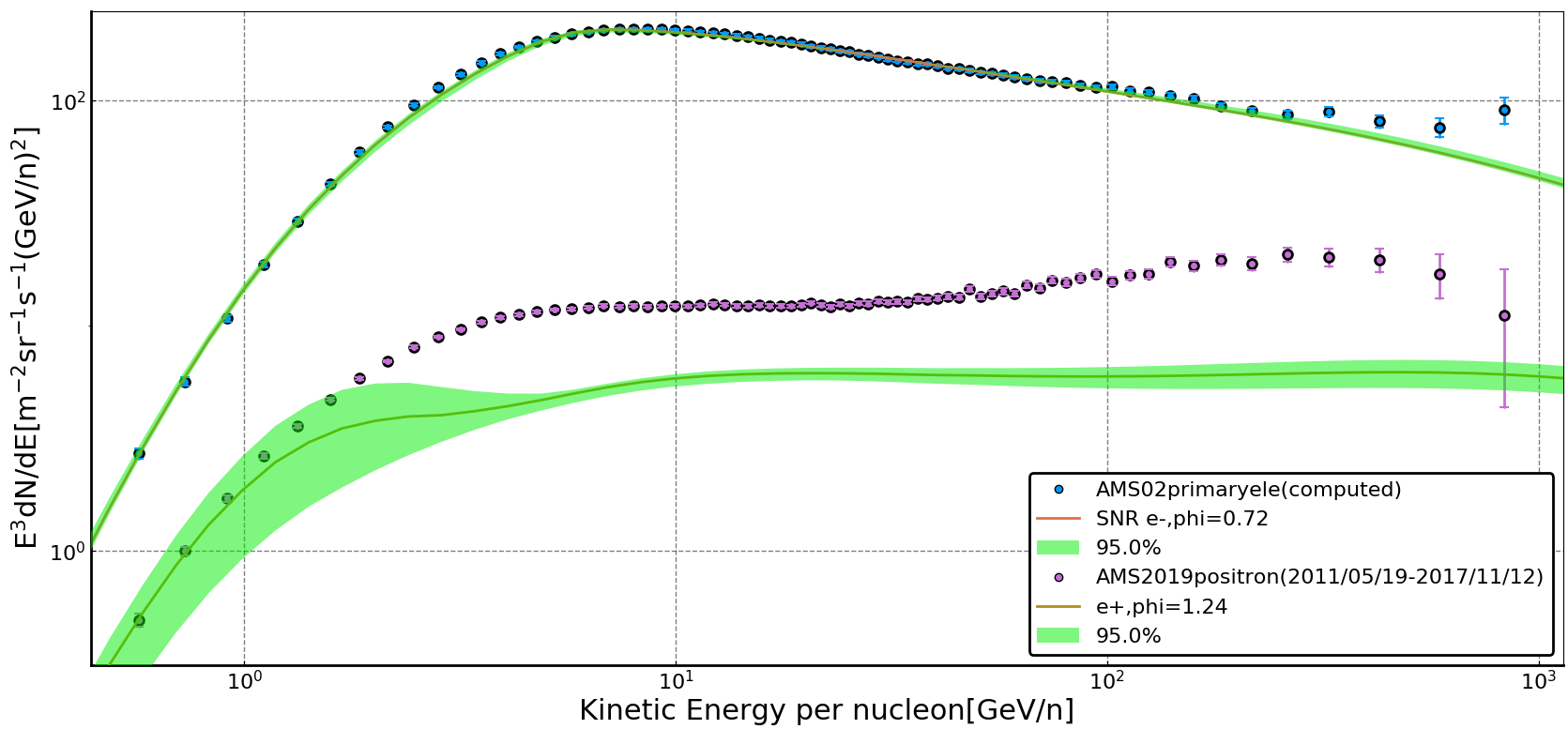}\\
\caption{Positron and primary electron spectra predicted by the best-fit parameters obtained in Sec.~\ref{sec:result} (green line) and the $2\sigma$ confidence interval (green band), compared with the experimental data of AMS-02 \cite{Aguilar:2021tos}. \label{fig:2sigma-e}}
\end{figure}

In Fig.~\ref{fig:2sigma-e}, we can see that the primary electron spectrum fits the data well below 100~GeV/n. As predicted by Eq.~(\ref{eq:pri}), a hard component at high energies will also give rise to electron spectral hardening, but the energy losses in the high-energy region are even stronger and steepen the electron spectrum so rapidly, making the result failed to reproduce the AMS-02 measurement above 100~GeV/n. Extra components above 100~GeV/n may be needed and may be located within relatively short distances, as electrons have a much shorter lifetime with strong energy losses. Young and nearby SNRs may produce a much harder component that gives rise to the excess \cite{Fang:2016wid,Fang:2017nww}.

We also find that a solar modulation potential larger than 1~GV is required for positrons to fit the data, which has also been noted by Ref.~\cite{Orlando:2017mvd}. The positron spectrum predicted by the SDD model is harder than that from standard models since we have introduced a hard component similar to the antiproton spectrum analyzed in Sec.~\ref{sec:pbar}, but the overall flux is still significantly lower than the AMS-02 measurement. The missing flux may come from some extra contribution of positrons sources, including nearby pulsars (or PWNe) or DM particles.

\subsection{Anisotropy}
In the diffusion approximation, the anisotropy is dominated by the radial streaming of the CR fluxes, and its amplitude $\hat{A}$ is computed as
\begin{equation}
\label{eq:aniso}
  \hat{A}=\frac{3D\left|\nabla\psi\right|}{v\psi}\propto\frac{D_i}{\Lambda(\rho)}({\rm local})\,.
\end{equation}
We note that the anisotropy amplitude has an anti-correlation with the B/C ratio (Eq.~(\ref{eq:s/p})). As the B/C ratio gets harder at higher energies, the anisotropy amplitude gets softer and its index would change from $\delta$ to nearly 0, which is different from standard models that assume an unchanged slope index.

\begin{figure}[htbp]
\includegraphics[width=0.5\textwidth]{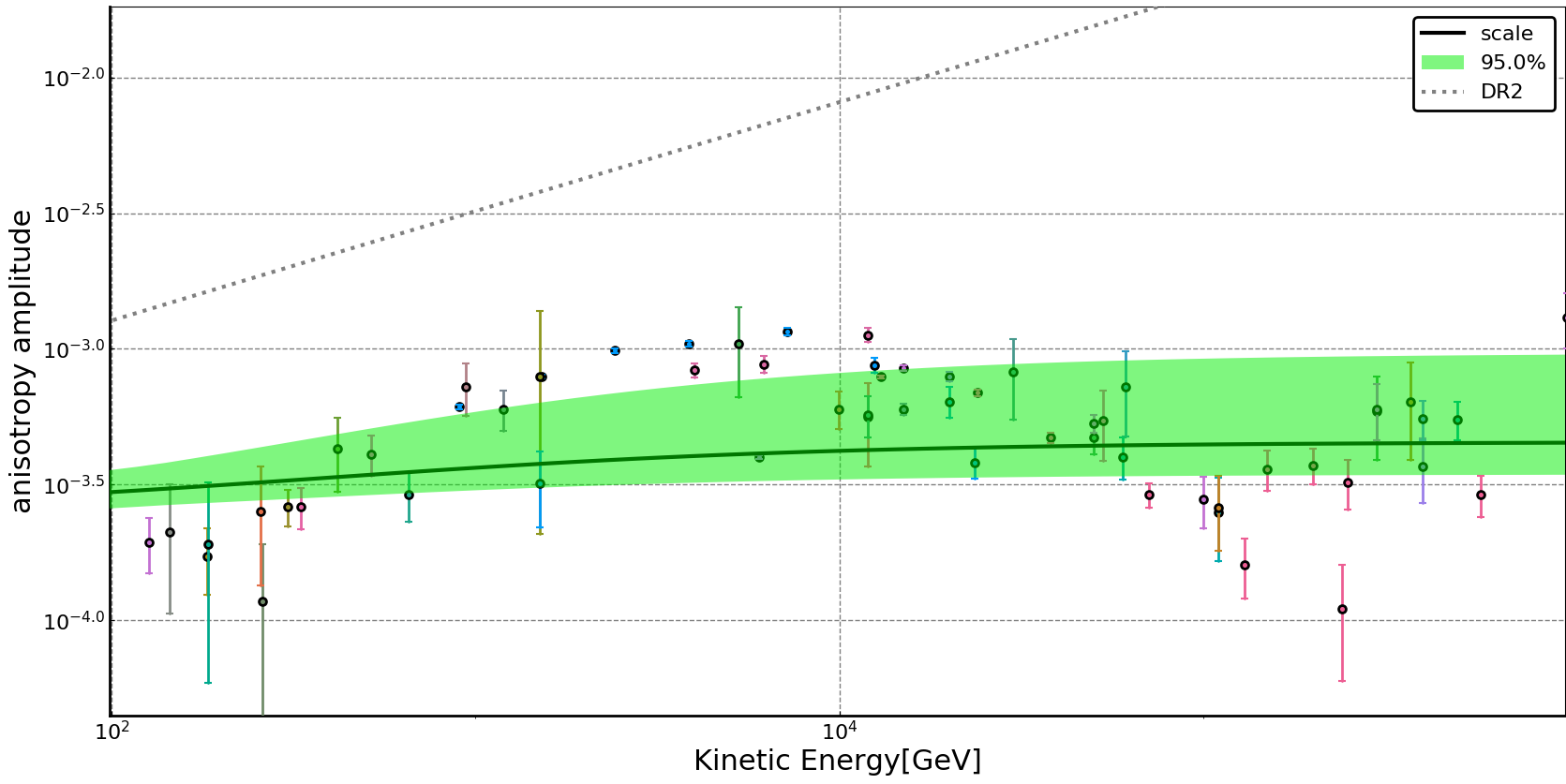}\\
\caption{\label{fig:2sig-ani} CR anisotropy amplitude predicted by best-fit parameters obtained in Sec.~\ref{sec:result} (black line) and $2\sigma$ confidence interval (green band), compared with the experimental data given in Ref.~\cite{Ahlers:2016rox}. An example of standard propagation model calculations (grey dotted line) are shown for reference \cite{Yuan:2018vgk}.}
\end{figure}

In Fig.~\ref{fig:2sig-ani}, we have drawn the anisotropy amplitude together with the uncertainties given by the fitting procedure in Sec.~\ref{sec:result}, which is consistent with the current observations (see Ref.~\cite{Ahlers:2016rox} and references therein). The grey dotted line is the anisotropy calculated with the standard propagation model, which is obviously higher than the observations. Besides, Ref.~\cite{Tomassetti:2012ga} pointed out that the anisotropy may be reduced in all energies if one accounts for a proper radial dependence for the diffusion coefficient.

The fine structures such as a sudden decrease of the anisotropy amplitude around $10^5$GeV are hard to be explained by the SDD model. There are other possibilities to further explain these features, for example, the presence of nearby sources of CRs \cite{Yuan:2020ciu}.

\section{SUMMARY\label{sec:conclusion}}
According to the assumption of different diffusion environments in the disk and the halo, we assume an SDD model for Galactic CR propagation, which has two different diffusion regions. In the inner region, the diffusion is slow and the slope index equals zero, while in the outer region, the diffusion coefficient is similar to the standard models. The SDD model has the advantage of naturally producing the observed high-energy spectral hardening of both the primary and secondary CR spectra without assuming a high-energy spectral break in the injection spectra or introducing nearby sources. By fitting the latest precise measurement of CR spectra and ratios, the parameters of the SDD model can be constrained in turn.

We perform a full Bayesian analysis based on an MCMC sampling algorithm to get the posterior parameter distributions. We use the carbon data as the primary CR spectrum, B/C ratio as the primary-to-secondary ratio, and $\rm{^{10}Be/^{9}Be}$ ratio as the unstable-to-stable ratio. The Be/B ratio is also adopted, the low-energy part of which can make up for the $\rm{^{10}Be/^{9}Be}$ data (which is not precise enough) and help to break parameter degeneracies.

The fitting result shows that all the parameters are well constrained. Our main finding is the well-constrained thickness of the SD region,  $h=0.468_{-0.062}^{+0.047}$~kpc at 95\% confidence, which could be the first precise estimation on this parameter. The fitted size of this region is a bit larger than the typical height of the Galactic disk, implying the convection of turbulent energy in the direction vertical to the Galactic plane. Other propagation parameters are consistent with those obtained in standard models, such as the height and diffusion coefficient of the outer diffusive halo.

Based on the fitting result, we also predict the $\bar p/p$ ratio, $e^-e^+$ fluxes, and the amplitude of CR anisotropy. We find that the $\bar p/p$ ratio fits well with the AMS-02 data, and no extra component is needed at high energies. The primary $e^-$ flux above 100~GeV is lower than the observation, which is likely due to the spectral fluctuation from nearby SNRs. The predicted positron flux is significantly lower than the AMS-02 data in a wide energy range, so primary positron sources are required, such as pulsars or DM. The anisotropy amplitude predicted by the SDD model fits the experimental data well in general, which is a remarkable advantage compared with standard models.

Other indirect observations such as CR-induced diffuse $\gamma$-ray emission \cite{Ackermann:2012pya} and molecular clouds emission \cite{Tibaldo:2015ooa} could be used to study the spatial variation of the diffusion coefficient in the Galaxy. The most prominent difference between the SDD and standard model is the vertical distribution of CRs, as the former predicts a more rapid decrease of the CR density within the SD region. The CR vertical distribution can be traced by the $\gamma$-ray emission of intermediate-velocity clouds located at various distances away from the Galactic plane \cite{Tibaldo:2015ooa}. More precise measurements of the clouds in the future may give a crucial test to the spatially dependent propagation model.

\acknowledgments
This work is supported by the National Natural Science Foundation of China under
the grants No. U1738209 and No. U2031110.

\bibliography{apssamp}
\appendix
\section{The Be/B grammage\label{app:beb}}
To analyze how the Be/B ratio is related to transport parameters, we firstly apply resolution on boron and beryllium nucleons respectively as
$\rm{Be=^{10}Be+^{9}Be+^{7}Be}$ and $\rm{B=^{11}B+^{10}B}$.
The collision with gas $(p+^{11}\rm{B)\to(^{10}Be+^{9}Be+^{7}Be})$ and decay process $^{10}\rm{Be\to^{10}B}+e^-$ also make effects, though not as important as the main production $\rm{(C-N-O)\to (Be-B)}$. 

For a more comprehensive review, one can refer to  \cite{Maurin:2001sj,Evoli:2019iih}, following which we can found that stable elements ($^{10}\rm{B}:I_a$) with contribution from
unstable ones ($^{10}\rm{Be}:I_b$) has this form:
\begin{equation}\label{be}
\frac{I_a}{X}=\sum_{a'>a} \frac{I_{a'}\sigma_{a'\to a}}{m} +\frac{I_bV_c}{\mu v}[\Delta \coth\frac{V_c\Delta L}{2D}-\coth\frac{V_cL}{2D}]\,.
\end{equation}
With diffuse-dominated grammage $X=\mu vL/2D$, decay-dominated grammage $X_d=\mu v\tau/\sqrt{4D\tau}$ and $\Delta=\sqrt{1+4D/V_c^2\tau}$, when unstable isotopes decay on a timescale shorter than $4D/V_c^2$ (usually below 100GV), $V_c\Delta\to\sqrt{4D/\tau}$ ,$V_c\to0$ (since we assumed few or no galactic wind here), the second term on the right side of Eq.~(\ref{be}) becomes 
$I_b\sqrt{4D/\tau}/{\mu v}$ and the third term becomes $2I_bD/L\mu v$.
Now we can write all isotopes below:
\begin{equation}
\begin{cases}

{\rm^{9}Be,^{7}Be}: \frac{I_a}{X}=\sum_{a'>a} \frac{I_{a'}\sigma_{a'\to a}}{m}\\
{\rm^{10}Be}: \frac{I_a}{X_d}=\sum_{a'>a} \frac{I_{a'}\sigma_{a'\to a}}{m} (<100{\rm GeV})\\
{\rm^{11}B}: \frac{I_a}{X}+\frac{I_{a}\sigma_{a}}{m}=\sum_{a'>a} \frac{I_{a'}\sigma_{a'\to a}}{m}\\
{\rm^{10}B}: \frac{I_a}{X}=\sum_{a'>a} \frac{I_{a'}\sigma_{a'\to a}}{m} +\frac{I_b}{X_d}-\frac{I_b}{X} (<100{\rm GeV})

\end{cases}\\
\end{equation}
By combining them we can further calculate different ratios as
\begin{equation}\label{eq:beb3}
\begin{cases}

{\rm\frac{^{9}Be+^{7}Be}{^{11}B}}: X\propto\frac{L}{2D}\\
{\rm\frac{^{10}Be}{^{11}B}}: X_d\propto\frac{\tau}{\sqrt{4D\tau}} (<100{\rm GeV})\\
{\rm\frac{^{10}Be}{^{10}B}}: \cfrac{1}{
\frac{X\sum \cfrac{I_{a'}\sigma_{a'\to a}}{m}}{X_d\sum \cfrac{I_{a'}\sigma_{a'\to b}}{m}}+\cfrac{X}{X_d}-1}\propto\frac{X_d}{X}=\frac{\sqrt{D\tau}}{L} (<100{\rm GeV})
\end{cases}\\
\end{equation}

So the total ratio of Be/B should be a mixture of all ratio forms in Eq.~(\ref{eq:beb3}), which shows the B/C-like (L/D) feature at high energy, $\rm{^{10}Be/^{9}Be}$-like ($\sqrt{D}/L$ and $1/\sqrt{D}$) feature at low energy.
The important feature for breaking the degeneracy of $D-L$ is $\rm{^{10}Be/^{9}Be}$-like. As available $\rm{^{10}Be/^{9}Be}$ ratio measurements have large uncertainties, an introduction of precise Be/B 
ratio is preferred. It is emphasized in \cite{Maurin:2001sj} that the $\rm{^{10}Be\to^{10}B}+e^-$
channel contributes up to 10\% of total Boron flux and cannot be neglected, but $\rm{^{10}Be}$ fluxes make up only 10\% of total beryllium at low energy and there are cross-section uncertainties, which result in a complicated problem.
\section{Above 20~GV}
\subsection{specified $h$\label{app:1}}
Before fitting the free parameters using all measurements according to Section~\ref{sec:data}, we firstly estimate how the SDD model fits the spectral hardening at high energy, by using AMS-02 carbon and boron fluxes, together with the Be/B ratio. All experiment points are taken above 20~GV, where the influences of low energy power-law break, solar modulation, cross-section uncertainties and non-relativity effect should be lowest.

To make the fitting much simpler we fix $\phi=0.8 \rm{GV},\eta=-0.5$, taken from Yuan's paper \cite{Yuan:2018vgk} as a reference of standard models.
We fix $N=8$ to give a rapid smoothness from halo to disk, and choose the disk thickness among specified values $h=\{0.3,0.5,0.8,1.0\}$~kpc as we have predicted a strong degeneracy of $h/\xi$ from Eq.~(\ref{eq:pri}).
 So the free parameters are $\bm{\theta}=\{D_0,\delta,L,V_a,\xi,\xi_{\delta}
,A_c,\nu\}$, consist of 6 transport parameters and 2 injection parameters.
\begin{figure*}
\includegraphics[width=0.8\textwidth]{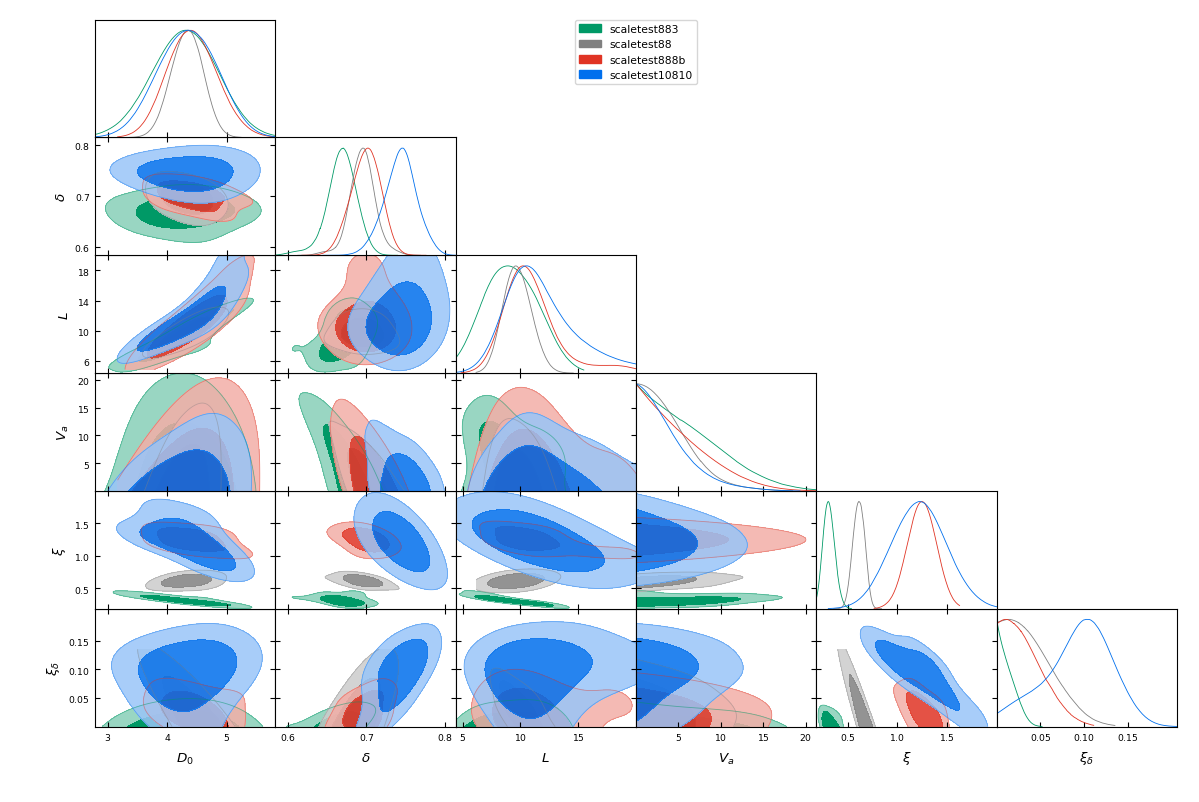}
\caption{\label{fig:wide}The 1-D and 2-D distributions of the transport parameters, different colors represent different values of specified $h$: 0.3~kpc(green), 0.5~kpc(grey), 0.8~kpc(red), 1.0~kpc(blue)}
\end{figure*}

\begin{figure}[htbp]
\includegraphics[width=0.5\textwidth]{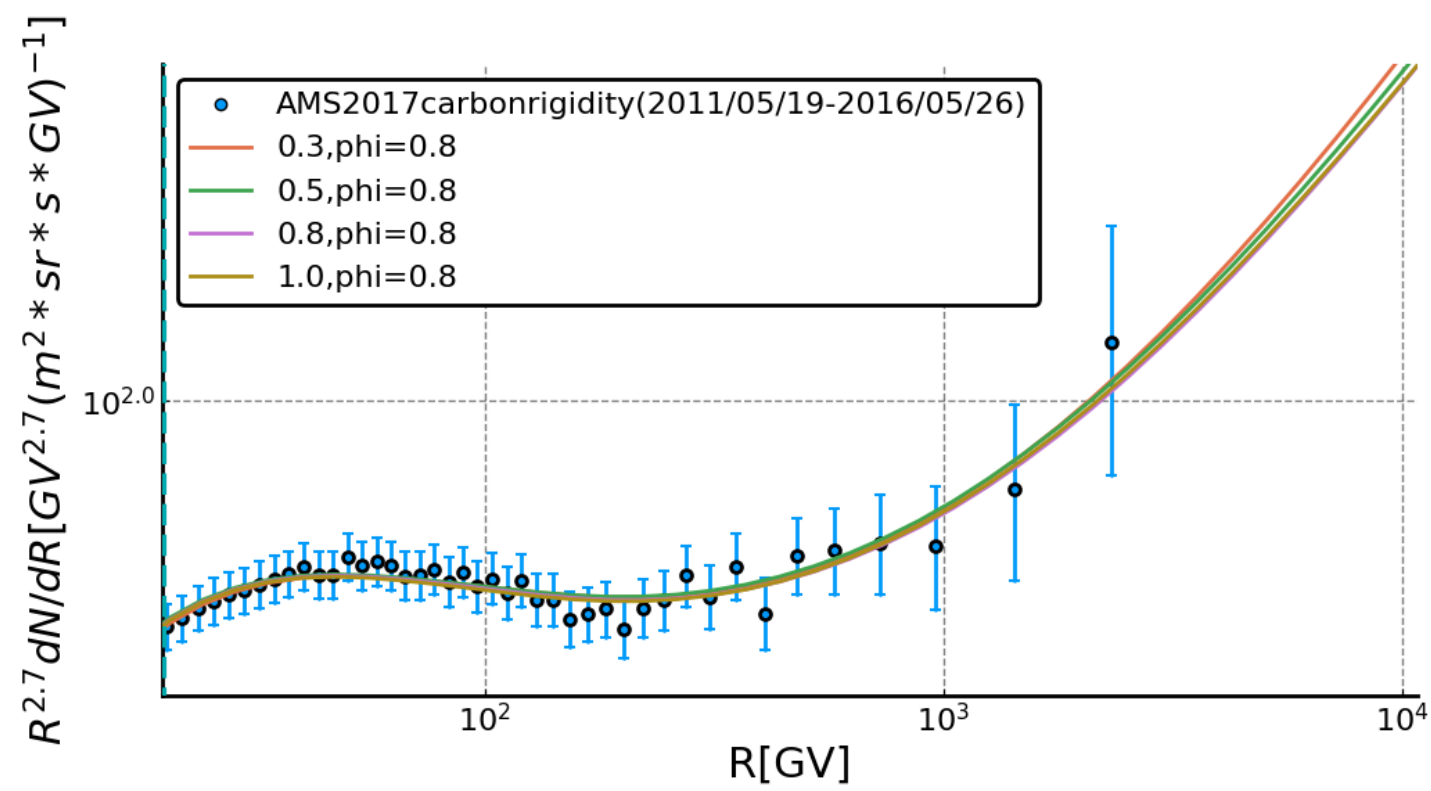}\\
\includegraphics[width=0.5\textwidth]{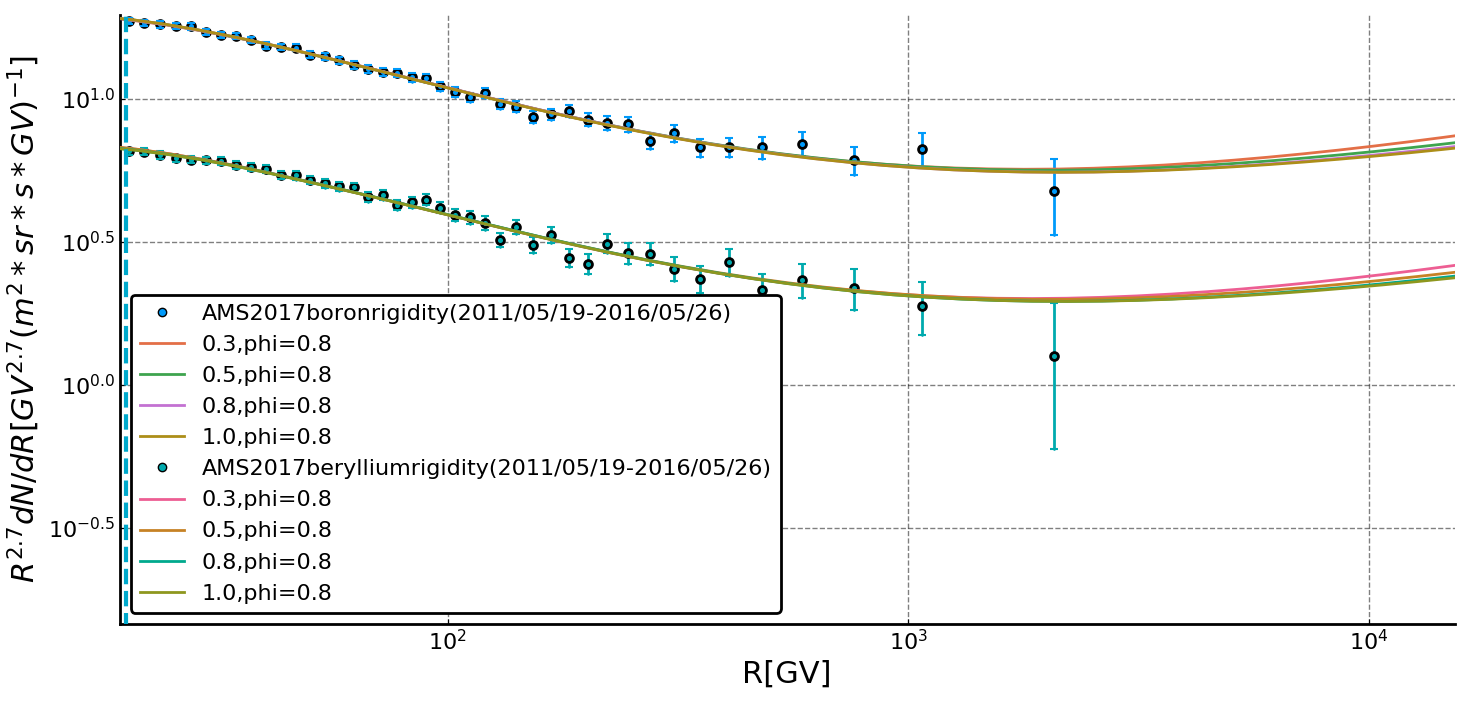}
\caption{Nucleon spectra calculated with the best-fit  parameters above 20~GV, compared with the experimental data AMS-02 \cite{Aguilar:2017hno,Aguilar:2018njt}, different colors represent different values of specified $h$. Top: carbon. Bottom: boron and beryllium. \label{fig:flux-h}}
\end{figure}
\begin{figure}[htbp]
\includegraphics[width=0.5\textwidth]{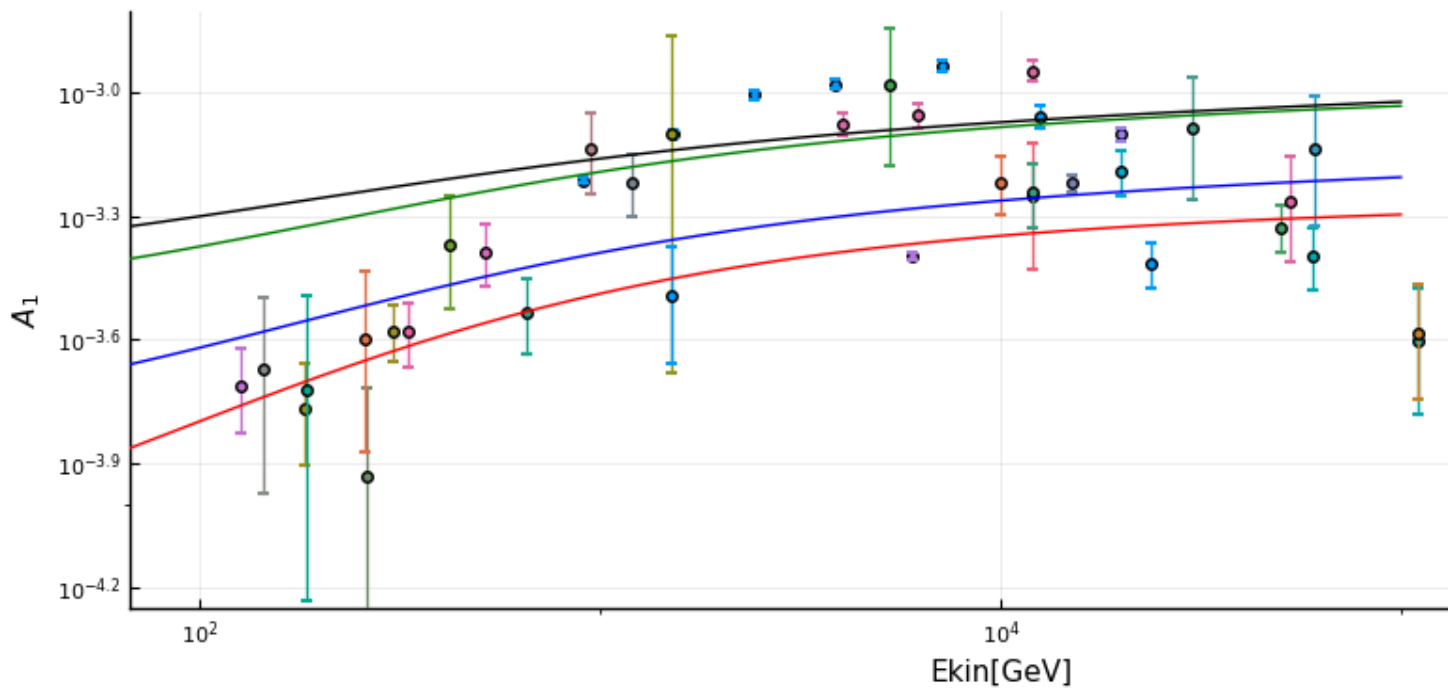}
\caption{CR anisotropy amplitude predicted by best-fit parameters, compared with the experimental data given in \cite{Ahlers:2016rox}, different colors represent different values of specified $h$: 0.3~kpc (red), 0.5~kpc (blue), 0.8~kpc (green), 1.0~kpc (black). \label{fig:ani-h}}
\end{figure}

\begin{table*}
\caption{\label{tab:tableh}The best-fit values and posterior 95\% range of all parameters in SDD model}
\begin{ruledtabular}
\begin{tabular}{ccccc}
 Parameter&h=0.3~kpc&h=0.5~kpc&h=0.8~kpc&h=1.0~kpc\\ \hline
 $D_0(10^{28}\rm{cm^2s^{-1}})$&4.648 [3.222,5.353]&4.451 [3.845,4.792]&4.076 [3.573,5.274]&3.823 [3.376,5.289]\\
 $\delta$&0.683 [0.631,0.707]&0.700 [0.662,0.729]&0.708 [0.663,0.733]&0.730 [0.701,0.784]\\
 $L$(kpc)&11.414 [4.747,13.703]&10.574 [7.418,12.144]&9.446 [6.458,18.076]&9.240 [6.387,19.176]\\
 $V_a$(km/s)&1.101 [0,13.409]&2.376 [0,8.862]&1.251 [0,12.969]&1.314 [0,9.701]\\
 $\xi$&0.2716 [0.1960,0.4251]&0.6181 [0.4953,0.7230]&1.3334 [0.9695,1.5242]&1.760 [0.6905,1.7528]\\
 $\xi_{\delta}$&0.002127 [0,0.03082]&0.02207 [0,0.08207]&0.02454 [0,0.06832]&0.03866 [0.01614,0.1612]\\
 $A_c(10^{-3})$\footnote{abundance of proton $A_p$ is $1.06*10^6$, and the normalization of proton flux at 100~GeV is $4.204*10^{-9} \rm{cm^{-2}s^{-1}sr^{-1}MeV^{-1}}$}&3.279 [3.243,3.319]&3.290 [3.254,3.321]&3.288 [3.259,3.335]&3.277 [3.249,3.325]\\
 $\nu$&2.352 [2.334,2.368]&2.359 [2.341,2.376]&2.363 [2.342,2.379]&2.360 [2.339,2.380]\\ \hline
 $\Delta$\footnote{$\Delta=\delta(1-\xi_{\delta})$}&0.651&0.662&0.669&0.673\\ \hline
 $\chi^2_{\rm{min}}/n_{\rm{dof}}$&47.51/118&47.29/118&46.78/118&47.13/118\\
\end{tabular}
\end{ruledtabular}
\end{table*}

Table~\ref{tab:tableh} and Fig.~\ref{fig:wide} shows the MCMC result of 4 kinds of specified $h$. We notice that all these $\chi^2_{\rm{min}}$s have similar values, thus the thickness $h$ does not show a strong preference to one of them. As half-thickness of inner disk gets larger, $\delta,\xi,\xi_{\delta}$ all get increased and a strong anti-correlation is shown between $\xi$ and $\xi_{\delta}$. It seems that the Alfv\'en velocity $V_a$ and slope index scale factor $\xi_{\delta}$ are all converged to zero. The effect of reacceleration does not significantly affect the hardening and could be ignored since we are now focused on the energy region above 20~GV. Moreover, the diffusion coefficient in the disk prefers an energy-independent type as $\xi_{\delta}\sim0$. In Fig.~\ref{fig:wide} we also find that the strong degeneracy of $D_0/L$ worsens the constraining of other parameters.

From Fig.~\ref{fig:flux-h} we find that the best-fit values of these specified $h$ have almost the same results fitting good with AMS-02 measurements, while a few dispersion appears above $10^4$~GV.
To estimate the hardening speed of the slow component $\rho^\Delta$, we calculate $\Delta$ in Table~\ref{tab:tableh} and find that this value keeps nearly unchanged when $h$ become larger, proving that $\Delta$ should be important in reproducing similar hardening features.

To estimate more differences of these specified $h$, we further predict the anisotropy amplitude in Fig.~\ref{fig:ani-h}. As $h$ gets larger, the amplitude in the entire energy range gets
larger (except 1.0~kpc), and the disk thickness $h\sim0.5$~kpc fits best with these experiments. Eq.~(\ref{eq:aniso}) has predicted a simple relation that $\hat A\sim \xi D_0/h=\{4.207,5.502,6.793,6.728\}$, which could explain those features.
\subsection{specified smooth factor $N$\label{app:2}}
\begin{figure}[htbp]
\includegraphics[width=0.5\textwidth]{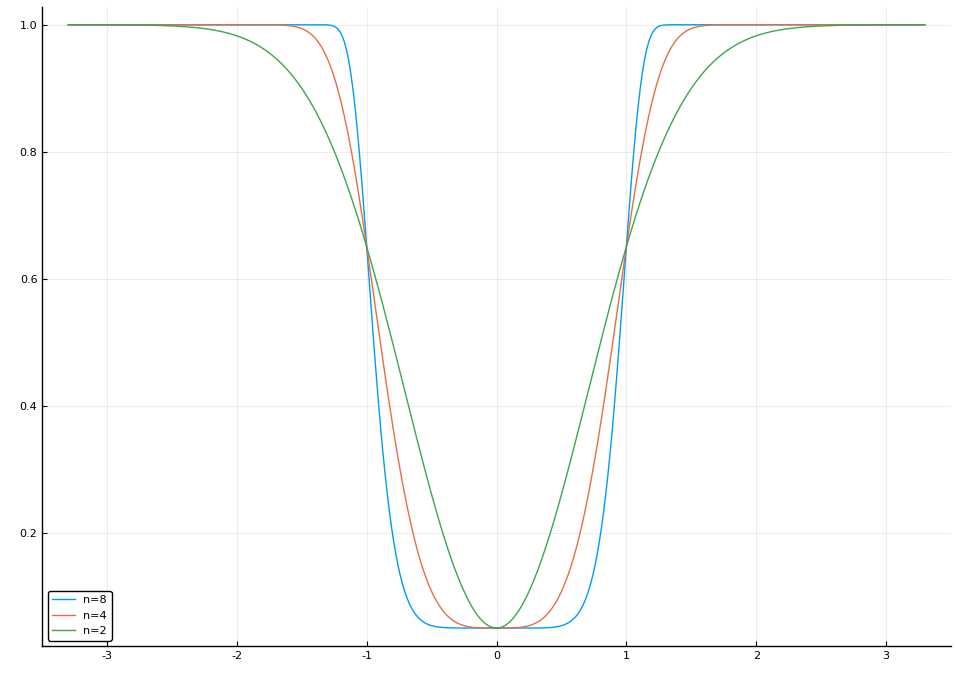}
\caption{The diffusion coefficient changes with different smooth factor $N$, where we assume $h=1$~kpc, $\xi=0.05$. \label{fig:N-D} }
\end{figure}

We have introduced the smooth factor $N$ in Eq.~(\ref{eq:diff}), which is used to describe the change of diffusion coefficient from the innermost disk to the outer halo. In Fig.~\ref{fig:N-D} we show the spatial dependence of diffusion coefficient, which changes more rapidly when smooth factor $N$ gets larger. When $N\sim\infty$, this change becomes a step function around the thickness $\pm h$ as Ref.~\cite{Tomassetti:2012ga} used this kind of spatial dependent model.

We choose $h=1.0$~kpc to give a fixed thickness of slow region and choose the smooth factor among specified values $N=\{2,4,8\}$. Other data sets are all referred to from Appendix.~\ref{app:1}.
Fig.~\ref{mcmc-n} shows the fitting results of 3 kinds of specified $N$. As smooth factor get larger, $\xi,\xi_{\delta}$ all get increased and the anti-correlation also appears between $\xi$ and $\xi_{\delta}$. The serious degeneracy of $D_0/L$ still worsens the constraining of other parameters.

\begin{figure}[htbp]
\includegraphics[width=0.5\textwidth]{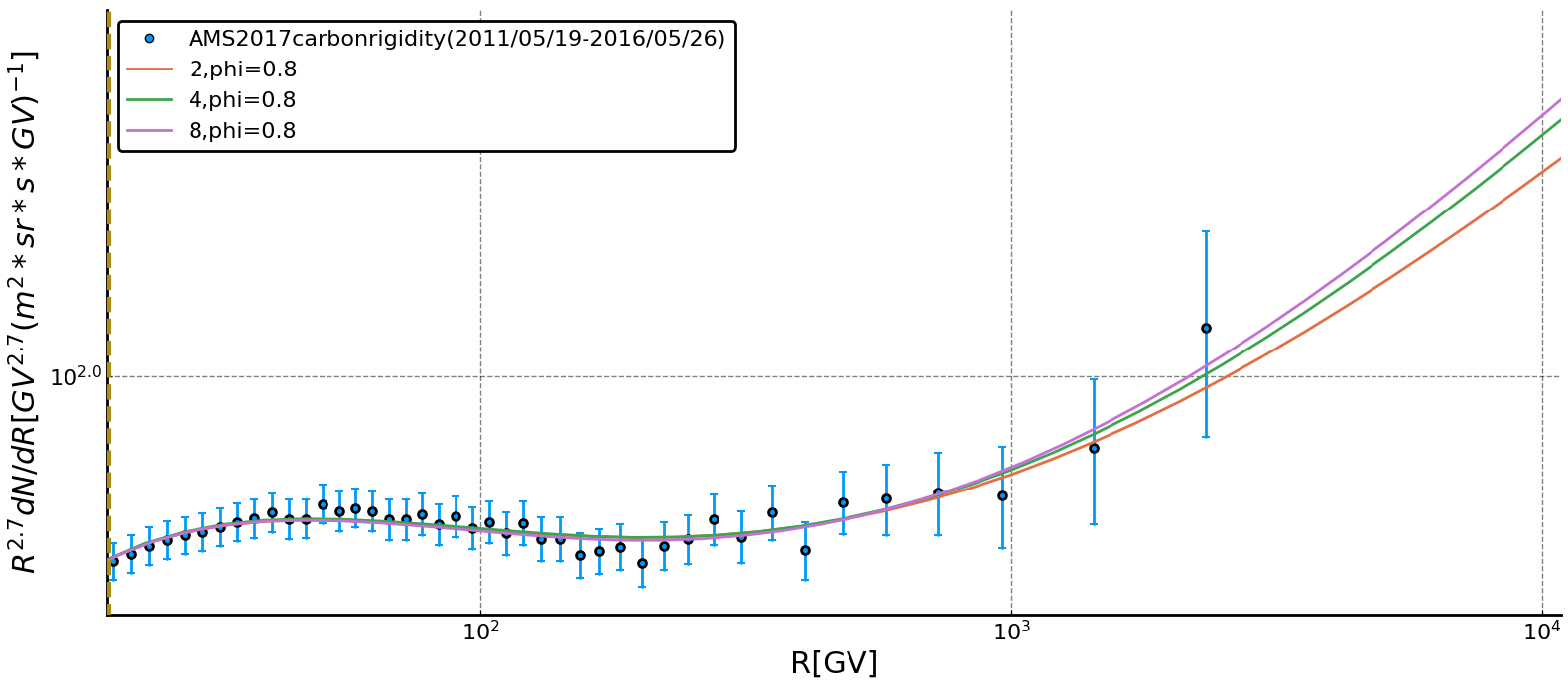}\\
\caption{Carbon spectrum calculated with the best-fit  parameters, compared with the experimental data AMS-02 \cite{Aguilar:2017hno} above 20~GV. Different colors represent different values of specified $N$. \label{fig:flux-N}}
\end{figure}

From Fig.~\ref{fig:flux-N} we find that the best-fit values of these specified $N$ have almost the same results fitting good with AMS-02 measurements, while large smooth factor $N$ can give stronger hardening above $10^3$~GV. The local CRs mainly travel from disk to halo and diffuse backward to the solar system. The real spectra may not be just the superposition of two components like Eq.~(\ref{eq:pri}) and Eq.~(\ref{eq:s/p}) but include more intermediate states.
 The presence of intermediate components can be used to explain how the hardening changes with $N$.

\begin{figure}[htbp]
\includegraphics[width=0.5\textwidth]{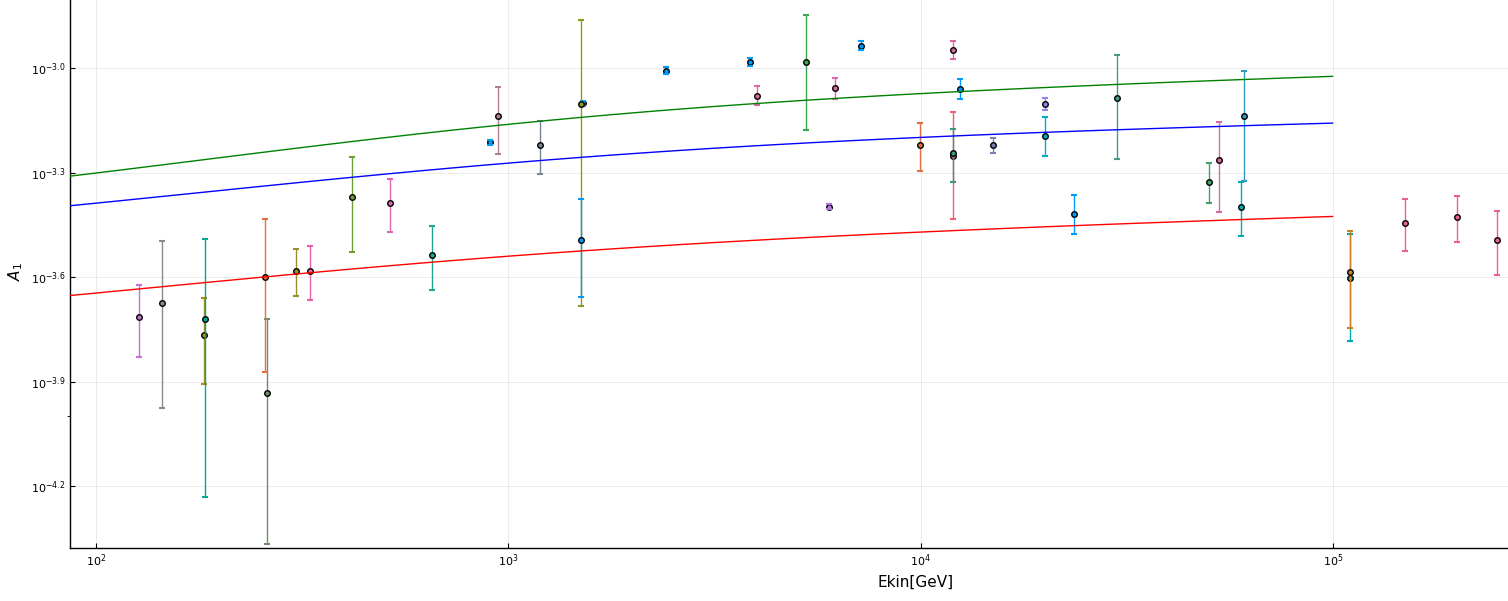}
\caption{CR anisotropy amplitude predicted by best-fit parameters, compared with the experimental data given in \cite{Ahlers:2016rox}. Different colors represent different values of specified $N$: 2(red), 4(blue), 8(green). \label{fig:ani-n}}
\end{figure}

\begin{figure*}
\includegraphics[width=0.8\textwidth]{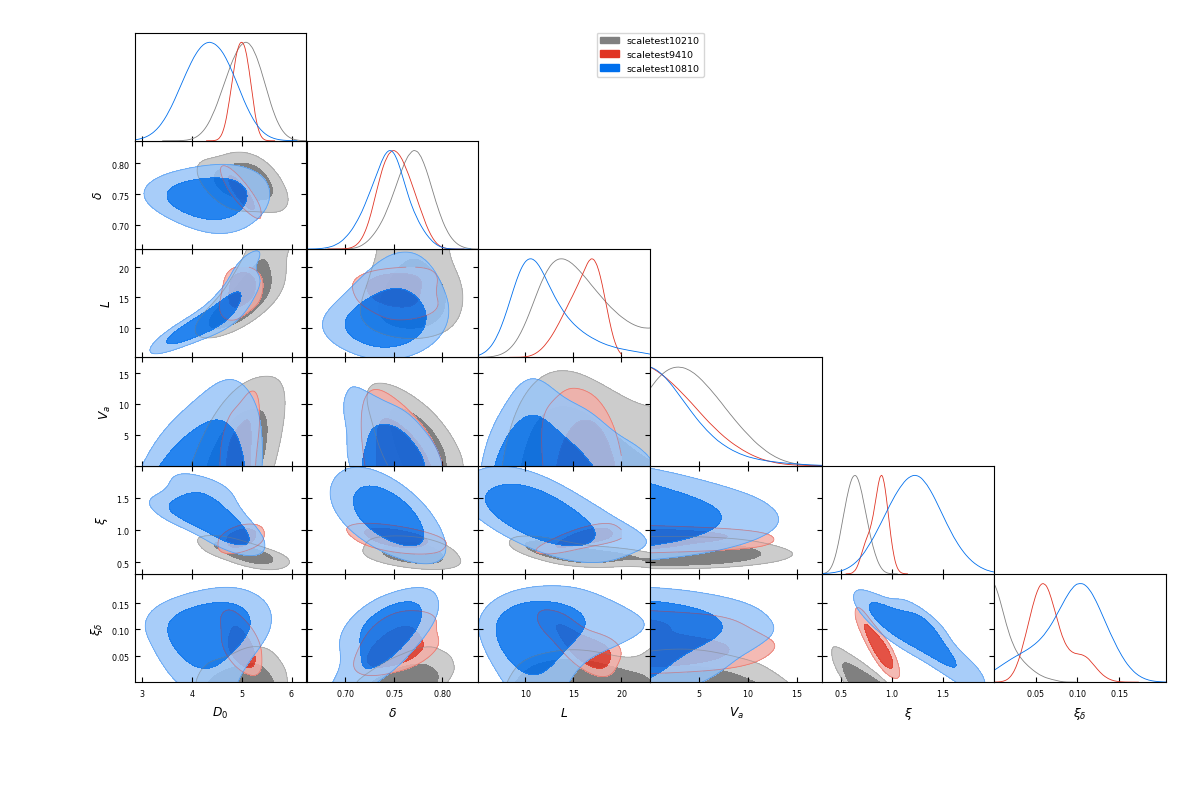}
\caption{\label{mcmc-n}The 1-D and 2-D distributions of the transport parameters. Different colors represent different values of specified $N$: 2(grey), 4(red), 8(blue)}
\end{figure*}

To estimate more differences of these specified $N$, we further predicted the anisotropy amplitude in Fig.~\ref{fig:ani-n}. As $N$ gets larger, the amplitude in the entire energy range gets
larger as well, and the smooth factor $N\sim4$ fits best with these experiments. Eq.~(\ref{eq:aniso}) has predicted a simple relation that $\hat A\sim \xi D_0/h=\{3.216,5.737,6.728\}$, which could explain those features shown in Fig.~\ref{fig:ani-n}.


\end{document}